\documentclass[prd,nofootinbib,showpacs,preprint]{revtex4}
\usepackage{graphicx}
\usepackage{epsfig}
\usepackage{epstopdf}
\usepackage{amsmath}
\usepackage{amsfonts}
\usepackage{amssymb}
\usepackage{url}
\usepackage{hyperref}
\usepackage{subfigure}
\usepackage{cancel}

\newcommand{\bd}{\begin{document}}
\newcommand{\ed}{\end{document}}
\newcommand{\bc}{\begin{center}}
\newcommand{\ec}{\end{center}}
\newcommand{\beqa}{\begin{eqnarray}} 
\newcommand{\eeqa}{\end{eqnarray}} 
\newcommand{\beq}{\begin{equation}} 
\newcommand{\eeq}{\end{equation}} 
\newcommand{\lsim}{\lesssim}
\newcommand{\gsim}{\gtrsim}
\newcommand{\nn}{\nonumber}
\newcommand{\bmt}{\begin{pmatrix}}
\newcommand{\emt}{\end{pmatrix}}
\usepackage[toc,page]{appendix}
\usepackage{comment} 
\def\TeV{\mbox{TeV}}
\def\GeV{\mbox{GeV}}
\def\MeV{\mbox{MeV}}
\def\eV{\mbox{eV}}
\def\keV{\mbox{keV}}
\def\s1{\hat s}
\def\ds{\displaystyle}
\def\lb{\Lambda_b}
\def\ll{\Lambda}
\newcommand{\be}{\begin{equation}}
\newcommand{\ee}{\end{equation}}
\newcommand{\bea}{\begin{eqnarray}}
\newcommand{\eea}{\end{eqnarray}}
\newcommand{\bref}[1]{(\ref{#1})}
\def\slashi#1{\rlap{\sl/}#1}
\newcommand{\<}[1]{\langle {#1} \rangle}
\begin{document}
\title{Effects of scalar leptoquark on semileptonic $\Lambda_b$ decays}
\author{Suchismita Sahoo and Rukmani Mohanta }
%\email{suchismita@uohyd.ac.in,  rmsp@uohyd.ac.in}
\affiliation{\,School of Physics, University of Hyderabad, 
              Hyderabad - 500046, India  }      
%%%%%%%%%%%%%%%%%%%%%%%%%%%%%%%%%%%%%%%%%%%%%%%%%%%%%%%%%%%%
\begin{abstract}

We study the  scalar leptoquark effects on the rare semileptonic  decays of $\Lambda_b$ baryon, governed by the quark level transition $b \to s l^+ l^-$. We estimate the branching ratios,  forward-backward asymmetries,  lepton polarization  parameters and the lepton flavour non-universality
effects   in these decay channels. We find significant deviations from the
corresponding standard model predictions in some of the observables due to leptoquark effects. We  also investigate the  lepton flavour violating decays $\Lambda_b \to \Lambda l_i^- l_j^+$, the branching ratios of which are found to be ${\cal O}(10^{-10} - 10^{-9})$.

\end{abstract}
\pacs{13.30.Ce, 14.80.Sv}

\maketitle

\section{Introduction}
The study of the rare $B$ meson decays involving flavour changing neutral current (FCNC) transitions  is very crucial, as they provide sensitive probe to look for new physics (NP) beyond the standard model (SM). These decays  are highly suppressed in the SM due to Glashow-Iliopoulos-Maiani (GIM) mechanism and occur only through one-loop level penguin and box diagrams.
Recently, several anomalies have been observed in the rare semileptonic $B$ decays  mediated through the FCNC $b \to s$ transitions. The most prominent ones are the observation of $3.7\sigma$ deviation in the angular observable $P_5^\prime$ \cite{lhcb1, lhcb1a, p5p} of $B \to K^* \mu^+ \mu^-$ mode 
and the violation of lepton universality in the $B \to K l^+ l^-$ decays at the level of $2.6 \sigma$  \cite{lhcb2} by the LHCb experiment. In addition, LHCb has also observed significant discrepancy in the decay 
rates of the $B \to K^* l^+ l^-$ processes \cite{lhcb3,  lhcb4}. Also the decay rate of  the $B_s \to \phi \mu^+ \mu^-$ process \cite{lhcb5} has $3.3\sigma$ deviation form its SM value in the low $q^2$ region.  Furthermore, the observed discrepancy in the ratio of branching fractions of exclusive $B \to K^{(*)} l^+ l^-$ decay and the inclusive decays into dimuon over dielectron in the full $q^2$ range \cite{pdg} provide strong evidence of the presence of lepton non-universality.

The anomalies observed in $b \to s l^+ l^-$ processes at LHCb \cite{lhcb1, lhcb1a, lhcb2, lhcb3, lhcb4, lhcb5} have attracted a lot of attention in recent times. The implications of these observations have been extensively studied both in the context of various new physics models and in model independent ways \cite{matias1, jager, jager15, huber, beaujean}. These deviations
which are at the level of (2-3)$\sigma$ are not statistically significant enough to provide an unambiguous   signal of new physics.  On the other hand they are also not small enough to  be ignored completely and need to  be scrutinized meticulously as many different ways as possible. If indeed they really evince the smoking gun
signal of some kind of NP, such effects must also show up in other
decay channels involving $b \to s$ transitions, such as 
the corresponding $\Lambda_b$ transitions. Therefore, the study of the rare $\Lambda_b$ decays is of utmost
importance to obtain an unambiguous signal of new physics. Including
the baryonic decay mode $\Lambda_b \to \Lambda(\to p \pi^-) \mu^+ \mu^-$
in the Bayesian analysis of $|\Delta B|=|\Delta S|=1$ transitions, a fit of the Wilson coefficients $C_{9,10,}$, $C'_{9,10}$ has been performed in Ref. \cite{dyk}, and it has been shown that, the shift to $C_9$ prefers to be 
opposite to the one found in mesonic case. To be more specific, the shift in $C_9$ in baryonic decay is found to be $\Delta_9=C_9-C_9^{SM}=1.6_{-0.9}^{+0.7}$~, as compared to the mesonic case where
its value is $\Delta_9=-1.09_{-0.20}^{+0.22}$ \cite{jager15}. Whereas the corresponding shifts in $C_{10}$ are in the same direction, i.e.,
$\Delta_{10}=0.7_{-0.8}^{+0.5}$ for the baryonic case $\Delta_{10}=0.56_{-0.24}^{+0.25}$ for mesonic case.  
%%%%%%%%%%%%%%%%%%%%%%%%%%%%%%%%%%%%%%%%%%%%%%%%%%%%%%%%%
As pointed out in \cite{dyk}, the observed discrepancy in the shift of $C_9$ might arise from our incomplete
understanding of the hadronic matrix elements of
the two-point correlators of ${\cal O}_{1,\cdots, 6;8}$  with the quark
electromagnetic current, which effectively shift the
Wilson coefficients $C_7$ and $C_9$. This could also be due to the large experimental uncertainties for the
$\Lambda_b \to  \Lambda (\to p \pi) \mu^+ \mu^-$  observables. 
However, if this persists with improved statistics,  this would 
constitute a  breakdown of the universal structure of the
transversity amplitudes at low recoil, as predicted by the operator product expansion (OPE).

%%%%%%%%%%%%%%%%%%%%%%%%%%%%%%%%%%%%%%%%%%%%%%%%%%%%%%%%%%

 The important distinction between the $\Lambda_b$  baryon and $B$ meson decays is the spin of the $\Lambda_b$ baryon.  Therefore, the number of degrees of freedom involved in the bound state of baryon is more, hence the systematic study of $\Lambda_b \to \Lambda \gamma$ and $\Lambda_b \to \Lambda \mu^+ \mu^-$ are relatively less explored in comparison to their mesonic counter parts. Also the experimental data on various $\Lambda_b$
 decay channels are rather limited. Recently LHCb has  reported the branching ratio of $\Lambda_b \to \Lambda \mu^+ \mu^-$ \cite{lhcb-lambda}, which is found to be lower than its standard model prediction. This decay process has been extensively studied 
in the literature both in the SM and in  various beyond the SM scenarios
\cite{Azizi, LCSR, lambda, Lattice-QCD, mohanta3, chen1, chen2, Latt-QCD-2}. To supplement these studies, in this paper  we would like to analyze
the rare baryonic decay processes $\Lambda_b \to \Lambda l^+ l^-$, where
$l=e,\mu,\tau$ in the scalar leptoquark model. In recent times, the scalar leptoquark model has been received a lot of attention, as it can successfully explain  most of the  observed anomalies associated with the $b \to s ll$ transitions.
 Leptoquarks are color-triplet bosonic particles
which can couple to a
quark and  lepton pair at the same time.
 The existence of leptoquark has been proposed in many extensions of the SM, such as grand unification model \cite{georgi, georgi2}, Pati-Salam model \cite{pati},  extended technicolor model \cite{schrempp} and the composite models \cite{kaplan}. The leptoquark states can be classified as vectors (spin-1) or scalars (spin-0). They can be characterized by their Fermion no.
$F=3B+L$, where $B$ and $L$ are the baryon no. and lepton no. respectively. Scalar leptoquarks may exist at TeV scale, and can give observable signatures in various low energy processes \cite{arnold}.
  The phenomenology of scalar leptoquarks  has been studied extensively in the literature \cite{mohanta1, mohanta2, davidson, arnold, kosnik, leptoquark, mohanta4, mohanta-Bs-mixing, wang}.
In this paper, we would like to study  the  rare  baryonic decay processes $\Lambda_b \to \Lambda l^+ l^-$ in the scalar leptoquark model. In particular, we estimate the decay rates, forward-backward  $(A_{FB})$ and  lepton polarization asymmetries  in these modes. 
 Furthermore,  we  explore the possibility of lepton non-universality 
 parameter in $\Lambda_b$ decays and also  the   lepton flavour violating (LFV)   decays mediated via the scalar leptoquarks.

The   paper is organized as follows. In Section II we present  the effective Hamiltonian responsible for the  $b \rightarrow s l^+ l^-$ processes and the decay parameters for the semileptonic $\Lambda_b \to \Lambda l^+ l^-$ decays in the standard model.  The new physics contribution due to the exchange of scalar leptoquark has been presented in section III   and  the constraints on the leptoquark parameter space
has been obtained  by using the  measured branching ratios of the rare decays $B_s \rightarrow l^+ l^- $. In section IV, we present the numerical analysis for the branching ratios and other physical observables such as the forward-backward asymmetry,  lepton polarization asymmetry and the lepton non-universality by using the constrained leptoquark couplings. We compute the   branching ratios of the lepton flavour violating $\Lambda_b \to \Lambda l_i^- l_j^+$  decays in section V and section VI contains the summary and  conclusion.

%%%%%%%%%%%%%%%%%%%%%%%%%%%%%%%%%%%%%%%%%%%%%%%%%%%%%%%%%%%%%%%%%%%%%%%%%%%%%%%
\section{Theoretical Framework  for the analysis of $\Lambda_b \to \Lambda l^+ l^-$ decay process } 
%%%%%%%%%%%%%%%%%%%%%%%%%%%%%%%%%%%%%%%%%%%%%%%%%%%%%%%%%%%%%%%%%%%%%%%%%

In this section, we will  discuss the SM contributions to the  branching ratios and other physical observables of the $\Lambda_b \to \Lambda l^+ l^-$, $l = e,\mu, \tau$ processes.
 The effective Hamiltonian describing the  decay process $\Lambda_b \to \Lambda l^+ l^-$ involves the quark level transition $ b \to s l^+ l^-$ and is given by \cite{buras} 
\bea
{\cal H}_{eff} &=& - \frac{ 4 G_F}{\sqrt 2} V_{tb} V_{ts}^* \Bigg[\sum_{i=1}^6 C_i(\mu) O_i +C_7 \frac{e}{16 \pi^2} \Big(\bar s \sigma_{\mu \nu}
(m_s P_L + m_b P_R ) b\Big) F^{\mu \nu} \nn\\
&&+C_9^{eff} \frac{\alpha}{4 \pi} (\bar s \gamma^\mu P_L b) \bar l \gamma_\mu l + C_{10} \frac{\alpha}{4 \pi} (\bar s \gamma^\mu P_L b)
\bar l \gamma_\mu \gamma_5 l\Bigg]\;,\label{ham}
\eea
where $V_{qq^\prime}$ are the CKM matrix elements, $G_F$ denotes the Fermi constant,  $\alpha$ is the fine-structure constant, $C_i$'s are the Wilson coefficients evaluated at the renormalized scale $\mu = m_b$ \cite{wilson} and $P_L,P_R = (1\mp \gamma_5)/2$ are the chiral operators. 
The sum over $i$ includes the current-current
operators $i = 1, 2$ and the QCD-penguin operators $i = 3, 4, 5, 6$.

In addition to the short distance contributions these processes also receive additional contributions arising from the long distance effects due to  the real $c \bar c$ resonant states of $J/\psi, \psi^\prime$, i.e.,
$\Lambda_b \to \Lambda J/\psi (\psi^\prime) \to \Lambda l^+ l^-$. These resonance contributions can be included by modifying the   Wilson coefficient $C_9$. Thus, the modified coefficient ($C_9^{eff}$) contains a perturbative part and a 
resonance part which can be  written  as 
\be
C_9^{eff}=C_9^{SM}+Y(s)+C_9^{res}\;,
\ee
where $C_9^{SM}$ is the SM Wilson coefficient evaluated at the $b$ quark mass scale \cite{wilson},  the perturbative part $Y(s)$ receives contributions coming 
from one-loop matrix elements  of the four quark operators \cite{buras2} and the long distance resonance effect is given by \cite{res}
\bea
C_9^{res}= \frac{3 \pi}{\alpha^2}(3 C_1+C_2+3C_3+C_4+3C_5+C_6)\sum_{
V_i=\psi(1S), \cdots, \psi(6S) } \kappa_{V_i}\frac{m_{V_i} \Gamma(V_i \to l^+ l^-)}{m_{V_i}^2 -s
-i m_{V_i}\Gamma_{V_i}}\;.
\eea
Here the phenomenological parameter $\kappa$ is taken to be $1.65$ and $2.36$ \cite{caso} for the lowest resonances $J/\psi$ and $\psi^\prime$ respectively in order to
reproduce the correct branching ratio of $ {\cal B}(B \to J/\psi K^*
\to K^* l^+ l^-)={\cal B}(B \to J/\psi K^*){\cal B}(J/\psi \to l^+ l^-)$.

The matrix elements of the  hadronic currents in (\ref{ham}) between initial $\Lambda_b$ and the
final $\Lambda$ baryon can be  parameterized in terms of various form factors which are presented in Appendix A.
Thus, using these matrix elements, the transition amplitude for the $\Lambda_b \to \Lambda l^+ l^-$ processes can be written as  \cite{mohanta3, chen1}
 \bea
{\cal M}(\lb \to \ll l^+ l^-) &=& \frac{G_F~ \alpha}{\sqrt 2 \pi}
V_{tb} V_{ts}^* \Biggr[ \bar l \gamma_\mu l \Big\{
\bar u_\ll \Big(\gamma^\mu (A_1 P_R+B_1 P_L)+ i \sigma^{\mu \nu} q_\nu
(A_2 P_R +B_2 P_L) \Big)u_{\lb} \Big\}\nn\\
&+&\bar l \gamma_\mu \gamma_5 l \Big\{
\bar u_\ll \Big(\gamma^\mu  (D_1 P_R+E_1 P_L)+ i \sigma^{\mu \nu} q_\nu
(D_2 P_R +E_2 P_L)\nn\\
&+& q^\mu(D_3 P_R +E_3 P_L) \Big)u_{\lb} \Big\}\Biggr]\;,\label{eq1}
\eea
where the parameters $A_i$, $B_i$, $D_j$ and $E_j$ with $i= 1,2$, $j=1,2,3$ are defined as
\begin{eqnarray}
A_{i} &=&C_{9}^{eff}\frac{f_{i}-g_{i}}{2}-\frac{2m_{b}}{q^{2}}%
C_{7}\frac{f_{i}^{T}{}+g_{i}^{T}}{2},  \nonumber \\
B_{i} &=&C_{9}^{eff}\frac{f_{i}+g_{i}}{2}-\frac{2m_{b}}{q^{2}}C_{7}\frac{%
f_{i}^{T}-g_{i}^{T}}{2},  \nonumber \\
D_{j} &=&C_{10}\frac{f_{j}-g_{j}}{2},
\nonumber \\
E_{j} &=&C_{10}\frac{f_{j}+g_{j}}{2}.
\label{ceff}
\end{eqnarray}
Using the transition amplitude (\ref{eq1}), the double differential decay rate is given by
\bea
\frac{d^2 \Gamma}{d\hat s~ dz}=\frac{G_F^2~ \alpha^2}{2^{12} \pi^5}~
|V_{tb} V_{ts}^*|^2~m_{\lb}~ v_l~ \lambda^{1/2}(1, r, \hat s)~
{\cal K}(\hat s, z)\;, \label{e2}
\eea
where
\bea
{\cal K}(\hat s, z)={\cal K}_0(\hat s)+z~ {\cal K}_1(\hat s)+z^2~{\cal K}_2(\hat s)\;,
\eea
 $\hat s = s/m_{\Lambda_b}^2$ and $z= \hat{p}_B\cdot\hat{p}_l^+ $ is the angle between the momenta of $\Lambda_b$ and $l^+$ in the dilepton invariant mass frame. 
The complete expressions for  ${\cal K}_0(\hat s)$, ${\cal K}_1(\hat s)$ and ${\cal K}_2(\hat s)$ are given in Appendix B.  Here $v_l=\sqrt{1 - (4m_l^2/q^2)}$ and  $\lambda (1, r, \hat s) = (1-r)^2-2 \hat s(1+r)+ \hat s^2$ is the triangle function with $r =m_\Lambda/m_{\Lambda_b}$. 
 The physical allowed range for $s\equiv q^2$ is
\be
4 m_l^2 \leq s\leq (m_{\lb}-m_\ll)^2\;.
\ee
Another interesting observable is the zero-crossing of the forward-backward asymmetry, wherein the position of the zero value of the  forward-backward asymmetry parameter  ($A_{FB}$) is very useful to look for the new physics signal.
The normalized forward-backward asymmetry is defined as
\bea
A_{FB}(\hat s)=\frac{\ds{\int_0^1 \frac{d^2 \Gamma}{d \s1 dz}dz-\int_{-1}^0 
\frac{d^2 \Gamma}{d \s1 dz}dz}}
{\ds{\int_0^1 \frac{d^2 \Gamma}{d \s1 dz}dz+\int_{-1}^0 
\frac{d^2 \Gamma}{d \s1 dz}dz}},
\eea
which can be simplified to  
\be
A_{FB}(\hat s)=\frac{{\cal K}_1(\hat s)}{{\cal K}_0(\hat s)+{\cal K}_2(\hat s)/3}\;.\label{fb}
\ee
The polarization asymmetries $P_i$ ($i=L, N, T$) are defined  as 
\be
P_i(\hat s)= \frac{\ds{\frac{d \Gamma}{d \hat s}(\hat \eta= \hat e_i)-
\frac{d \Gamma}{d \hat s}(\hat \eta= -\hat e_i)}}
{\ds{\frac{d \Gamma}{d \hat s}(\hat \eta= \hat e_i)+
\frac{d \Gamma}{d \hat s}(\hat \eta= -\hat e_i)}}\;,
\ee
where $\hat e_i$'s are the unit vectors along the longitudinal, normal and transverse components of the $l^+$ polarization and $\hat \eta $ is a unit vector, used to write the $l^+$ four-spin vector ($s_+$), along the $l^+$ spin in its rest frame as
\bea
s_+^0= \frac{\vec p_+ \cdot \hat \eta }{m_l}\;,~~~\vec s_+= 
\hat \eta + \frac{s_+^0}{E_{l^+}+m_l}\vec p_+\;.
\eea
Thus, the observables $P_L$, $P_T$ and $P_N$ correspond to longitudinal, transverse and normal polarization asymmetries respectively.  The observables  $P_L$ and $P_T$ are $P$-odd, $T$-even, while $P_N$ is $P$-even, $T$-odd and $CP$-odd.  The explicit expressions for forward-backward asymmetry and all the polarization parameters are taken from \cite{chen1,chen2,mohanta3}.

Another interesting observable is the lepton universality violation (LUV)
parameter, which has been recently observed  by the  LHCb collaboration in $B^+ \to K^+ l^+ l^-$ process and has $2.6 \sigma$ deviation from its SM predicted value \cite{lhcb4}. Analogously, we define the parameter $(R_\Lambda)$ as the ratio of branching fractions of $\Lambda_b \to \Lambda l^+ l^-$ into dimuon over dielectron as
\bea
R_{\Lambda} = \frac{{\rm Br}(\Lambda_b \to \Lambda \mu^+ \mu^-)}{{\rm Br}(\Lambda_b \to \Lambda e^+ e^-)}.
\eea
%%%%%%%%%%%%%%%%%%%%%%%%%%%%%%%%%%%%%%%%%%%%%%%%%%%%%%%%%%%%%%%%%%%%%%%%%%%%% 
\section{New physics contribution due to scalar leptoquark exchange}
%%%%%%%%%%%%%%%%%%%%%%%%%%%%%%%%%%%%%%%%%%%%%%%%%%%%%%%%%%%%%%%%%%%%%%
In this section we will consider the effect of scalar leptoquarks to the
$\Lambda_b \to \Lambda l^+ l^-$ decay processes. 
The exchange of leptoquarks will contribute additional operators to the SM effective Hamiltonian and thus, the various observables may deviate significantly  from their corresponding SM values. The  scalar leptoquark multiplets  with representations $X(3,2,7/6)$ and $X(3,2,1/6)$  under the SM gauge group $SU(3)_C \times SU(2)_L \times U(1)_Y$ conserve baryon and lepton numbers and don't allow proton decay. These baryon and lepton number conserving scalar leptoquarks  can have sizable Yukawa couplings and could be light enough to be accessible in accelerator searches. Thus, they could potentially contribute to the $b \to s l^+ l^- $ transitions and one can constrain the underlying couplings from experimental data on $B_s \to l^+ l^-$ processes as well as from $B_s - \bar{B}_s$ mixing.

The interaction  Lagrangian of the scalar leptoquarks $X = (3,2,7/6)$ with the SM bilinear fermions is given as \cite{arnold, kosnik}
\bea
{\cal L}= -\lambda_u^{ij}~ \bar u_{ R}^i X^T \epsilon L^j_L - \lambda_e^{ij}~ \bar e_{ R}^i X^\dagger  Q^j_L + h.c.,\label{lag}
\eea
where  $i, j$ are the generation indices, $X$ is the leptoquark doublet,  $Q_L$ ($L_L$) denotes the left handed quark (lepton) doublet, 
   the right handed up-type quark (charged lepton) singlet is represented by $u_R$ ($e_R$)  and $\epsilon = i\sigma_2$ is a $2 \times 2$ matrix.  The multiplets defined above are represented as
 \bea
X=\begin{pmatrix}
 V_\alpha \\
 Y_\alpha \\
 \end{pmatrix}, ~~~~~Q_L=\begin{pmatrix}
u_L \\
 d_L \\
 \end{pmatrix} ,~~~~~ {\rm and}~~~~~L_L=\begin{pmatrix}
\nu_L \\
 e_L \\
 \end{pmatrix} .
 \eea
 Now expanding the $SU(2)$ indices, the interaction Lagrangian (\ref{lag}) takes the form 
\bea
{\cal L}= -\lambda_u^{ij}~ \bar u_{\alpha R}^i ( V_\alpha e_L^j - Y_\alpha \nu_L^j )
-\lambda_e^{ij}~ \bar e_R^i \left (V_\alpha^\dagger u_{\alpha L}^j + Y_\alpha^\dagger d_{\alpha L}^j \right )+h.c.\;.\label{lepto}
\eea 
Thus, from Eq. (\ref{lepto}) one can obtain the  interaction Hamiltonian for  $b \to s l_i^+ l_i^-$ processes after performing the Fierz transformation as
\bea
{\cal H}_{LQ}= \frac{\lambda_e^{i3} {\lambda_e^{i2}}^*}{8 M_Y^2}[\bar s \gamma^\mu (1-\gamma_5) b][\bar l_i \gamma_\mu (1+\gamma_5) l_i]=\frac{\lambda_e^{i3} {\lambda_e^{i2}}^*}{4 M_Y^2}(O_9+O_{10})\;.\label{hamlq}
\eea
Comparing (\ref{hamlq}) with the corresponding SM effective Hamiltonian
(\ref{ham}), one can obtain the new Wilson coefficients as
 \bea
C_9^{NP} = C_{10}^{NP} = - \frac{ \pi}{2 \sqrt 2 G_F \alpha V_{tb} V_{ts}^* }\frac{\lambda_e^{i3}{ \lambda_e^{i2 }}^*}{
M_Y^2}\;.\label{c10np}
\eea
Similarly, the interaction Lagrangian due to the exchange of the scalar leptoquark $X = (3,2,1/6)$ is 
 \bea
{\cal L} = - \lambda_d^{ij}~ \bar d_{\alpha R}^i (V_\alpha e_L^j-Y_\alpha \nu_L^j) +h.c.\;,
\eea
which contributes to the  primed Wilson coefficients ($C^\prime_{9,10}$) corresponding to the  semileptonic electroweak penguin operators $\mathcal{O}^\prime_{9,10}$ (i.e., the right-handed counter parts of the SM operators $\mathcal{O}_{9,10}$) and are given as
\bea
C_9^{'NP } = - C_{10}^{'NP } = \frac{ \pi}{2 \sqrt 2 ~G_F \alpha V_{tb}V_{ts}^*} \frac{\lambda_s^{2i} {\lambda_b^{3i}}^*}{M_V^2}\;.\label{c10np1}
\eea
%\subsection{Constraint on the leptoquark parameters} 
Thus, from the above Eqs. (\ref{c10np}) and (\ref{c10np1}), one can find that there are four additional Wilson coefficients $C_{9,10}^{(\prime)NP}$, which will contribute  to the $b \to s l^+ l^-$ processes due to the scalar leptoquark exchange. Thus,  the modified parameters (\ref{ceff}) in the amplitude (\ref{eq1}),   become 
\begin{eqnarray}
A_{i} &=&C_{9}^{'NP}\frac{f_{i}+g_{i}}{2}+(C_{9}^{eff}+C_9^{NP})\frac{f_{i}-g_{i}}{2}-\frac{2m_{b}}{q^{2}}%
C_{7}^{SM}\frac{f_{i}^{T}{}+g_{i}^{T}}{2},  \nonumber \\
B_{i} &=&(C_{9}^{eff}+C_9^{NP})\frac{f_{i}+g_{i}}{2}-\frac{2m_{b}}{q^{2}}C_{7}^{SM}\frac{%
f_{i}^{T}-g_{i}^{T}}{2}+C_{9}^{'NP}\frac{f_{i}-g_{i}}{2},  \nonumber \\
D_{j} &=&C_{10}^{'NP}\frac{f_{j}+g_{j}}{2}+(C_{10}^{SM}+C_{10}^{NP})\frac{f_{j}-g_{j}}{2},
\nonumber \\
E_{j} &=&(C_{10}^{SM}+C_{10}^{NP})\frac{f_{j}+g_{j}}{2}+C_{10}^{'NP}\frac{f_{j}-g_{j}}{2}.
\label{ceff1}
\end{eqnarray}
Next, we have to find out the constraints on the leptoquark couplings to see how various observables behave in the LQ model.
 The detailed calculation of the constraint on the new leptoquark parameter space  has been presented in \cite{mohanta1, mohanta2, mohanta-Bs-mixing}, therefore here we will simply quote  the main results. We constrain the leptoquark coupling by comparing the theoretical \cite{bobeth1} and experimental \cite{cms, lhcb6, lhcb7}  branching ratios of $B_s \to l^+ l^-$   processes and  the $B_s-\bar B_s$ mixing data \cite{pdg}. For completeness, here we briefly outline the procedure for obtaining the constraints from $B_s \to \mu^+ \mu^-$ process and $B_s- \bar B_s$ mixing,  however, the technical details can be found in \cite{mohanta1, mohanta2, mohanta-Bs-mixing}. 
\subsection{Constraint from $B_s \to \mu^+ \mu^-$ process} 
 In the leptoquark model the branching ratio for the $B_s \to \mu^+ \mu^-$ mode can be given as
 \bea
{\rm Br}(B_s \to \mu^+ \mu^-)&=& \frac{G_F^2}{16 \pi^3} \tau_{B_s} \alpha^2 f_{B_s}^2 M_{B_s} m_{\mu}^2 |V_{tb} V_{ts}^*|^2
\left |C_{10}^{SM}+C_{10}^{NP}-C_{10}^{'NP}\right |^2 \sqrt{1- \frac{4 m_\mu^2}{M_{B_s}^2}}\nn\\
&=&
{\rm Br}^{SM}\left | 1+ \frac{C_{10}^{NP} - C_{10}^{' NP}}{C_{10}^{SM}} \right |^2
\equiv {\rm Br}^{SM}\left | 1+ r e^{i \phi^{ NP}} \right |^2\;,
\eea
where ${\rm Br}^{SM}$ is the SM branching ratio and   the parameters $r$ and $\phi^{NP}$ are defined as
\be
r e^{i \phi^{NP}}=\frac{C_{10}^{NP} - C_{10}^{' NP}}{C_{10}^{SM}}\;.
\ee
Now comparing the SM theoretical prediction of ${\rm Br}(B_s \to \mu^+ \mu^-)$ \cite{bobeth1} 
\bea
{\rm Br}(B_s \to \mu^+ \mu^-)|_{\rm SM}&=&\left (3.65 \pm 0.23 \right ) \times 10^{-9},
\eea
with the corresponding experimental value \cite{cms, lhcb6, lhcb7}
\bea
{\rm Br}(B_s \to \mu^+ \mu^-)=\left (2.9 \pm 0.7 \right ) \times 10^{-9},
\eea
and assuming that each
individual leptoquark contribution to the branching ratio does not exceed the experimental result, 
one can obtain
the bound on the new physics  parameters $r$ and $\phi^{NP}$. 
The allowed parameter space
in $r-\phi^{NP}$ plane which is compatible with the $1\sigma$ range of
the experimental data is 
 \bea
 0\leq r \leq 0.35\;, ~~~~{\rm with}~~~~\pi/2 \leq \phi^{NP} \leq 3 \pi/2\;.\label{r-bound1}
 \eea
 These bounds can be
  translated to obtain the bounds for the leptoquark couplings  as
 \bea
 0 \leq \frac{|\lambda_{ \mu}^{23} {\lambda_{ \mu}^{22}}^*|}{M_{Y}^2}
 = \frac{|\lambda_{s}^{22} {\lambda_{ b}^{32}}^*|}{M_{V}^2}\leq 5 \times 10^{-9} ~ {\rm GeV}^{-2}~~~~{\rm for}~~~~\pi/2 \leq \phi^{NP} \leq 3 \pi/2\;.
 \eea
 Similarly, one can obtain the  upper bound on the product of various combination of leptoquark couplings from $B_s \to l^+ l^-$ processes which are  presented in Table I.  Using the  bounds on  leptoquark couplings one can obtain the constraints on new Wilson coefficients using the eqns. (\ref{c10np}) and (\ref{c10np1}).

\begin{table}[htb]
\begin{center}
\caption{Constraints  on the leptoquark couplings obtained from various leptonic $B_{s} \to l^+ l^-$ decays \cite{mohanta2}, where $M_S$ denotes the mass of the scalar LQ.}
\vspace*{0.1 true in}
\begin{tabular}{|c|c|c|}
\hline
~~Decay Process~~ &~~ Couplings involved~~ & ~~Bound on the~~  \\ 
             &   &~ LQ couplings (${\rm GeV^{-2}}$)~  \\
\hline
$B_s \to \mu^\pm \mu^\mp $ &~ $\frac{|\lambda^{32} {\lambda^{22}}^*|}{M_S^2}$ ~& ~$ < 5.0 \times 10^{-9} $~\\

\hline

$B_s \to e^\pm e^\mp $ &~ $\frac{|\lambda^{31} {\lambda^{21}}^*|}{M_S^2}$ ~& ~$ < 2.54 \times 10^{-5} $~\\

\hline

$B_s \to \tau^\pm \tau^\mp $ &~ $\frac{|\lambda^{33} {\lambda^{23}}^*|}{M_S^2}$ ~& ~$ < 1.2 \times 10^{-8} $~\\
\hline
\end{tabular}
\end{center}
\end{table}
%%%%%%%%%%%%%%%%%%%%%%%%%%%%%%%%%%%%%%%%%%%%%%%%%%%%%%%
\subsection{Constraint from $B_s-\bar B_s$ mixing}
In this subsection, we will discuss the constraint on leptoquark couplings from  the $B_s-\bar B_s$ mixing, which in the SM, proceeds through the box diagram with internal top quark and $W$ boson exchange. The effective Hamiltonian describing the $\Delta B=2$
transition  is given as \cite{lim} 
\bea
 {\cal H}_{eff}=\frac{G_F^2}{16 \pi^2}~ |V_{tb} V_{ts}^*|^2~ M_W^2 S_0(x_t)\eta_B
(\bar s b)_{V-A}(\bar s b)_{V-A}\;, 
\eea 
where $\eta_B$ is the QCD correction factor and $S_0(x_t)$ is
the loop function given in Ref. \cite{lim}.
%\be S_0(x_t)=\frac{4 x_t -11 x_t^2
%+x_t^3}{4(1-x_t)^2} - \frac{3}{2} \frac{\log x_t x_t^3}{(1-x_t)^3}\;,
%\ee with $x_t=m_t^2/M_W^2$. 
Thus, the $B_s - \bar B_s$ mixing
amplitude in the SM, can be written as \be M_{12}^{SM}=\frac{1}{2
M_{B_s}} \langle \bar B_s|{\cal H}_{eff}| B_s \rangle = \frac{G_F^2}{12
\pi^2} M_W^2~ |V_{tb} V_{ts}^*|^2~ \eta_B~ \hat B_s f_{B_s}^2 M_{B_s} S_0(x_t)\;.
\label{sm} \ee 
 The corresponding mass difference can be computed from  the mixing amplitude through $\Delta M_s = 2 |M_{12}|$.
Now using the particle masses from \cite{pdg}, $\eta_B = 0.551$, the Bag parameter $\hat B_{B_s}=1.320 \pm 0.017 \pm 0.030$ and
the decay constant $f_{B_s}=225.6 \pm 1.1 \pm 5.4 $ from \cite{ckmfitter},  the value of
$\Delta M_s$ in the SM, is found as
\bea
\Delta M_s^{SM} = (17.426\pm 1.057)~ {\rm ps^{-1}},
\eea
which is in  good agreement with the experimental result \cite{pdg}
 \bea
\Delta M_s = 17.761 \pm 0.022~ {\rm ps^{-1}}.\label{mass-diff}
\eea
 For $X(3,2,7/6)$ LQ, the  mixing amplitude  receives additional contribution  from leptoquark and  charged lepton in the box diagram whereas for 
 $X(3,2,1/6)$ both charged lepton and neutrino will contribute to the mixing amplitude.  
The effective Hamiltonian due to the leptoquark $X(3,2,7/6)$  is given by
\be
{\cal H}_{eff}=\sum_{i=e,\mu,\tau} \frac{(\lambda^{bi} {\lambda^{si}}^{*})^{2}}{128 \pi^2}\frac{1}{ M_{S}^2}~I
\left (\frac{m_i^2}{M_S^2} \right )(\bar b \gamma^\mu P_L s) (\bar b \gamma_\mu P_L s)\;,
\ee 
and for  $X(3,2,1/6)$ leptoquark the corresponding effective Hamiltonian
becomes 
\be
{\cal H}_{eff}=\sum_{i=e,\mu,\tau} \frac{({\lambda^{bi}}^{*} \lambda^{si})^{2}}{128 \pi^2}\left [ \frac{1}{ M_{S}^2}~I
\left (\frac{m_i^2}{M_S^2} \right )+\frac{1}{ M_S^2} \right ] (\bar b \gamma^\mu P_R s) (\bar b \gamma_\mu P_R s)\;.
\ee 
where the loop function  $I(x)$ is given as 
\be
I(x)=\frac{1-x^2+2x \log x}{(1-x)^2}.
\ee
Thus, the contribution to the mixing amplitude due to the exchange of scalar leptoquark is given by
\bea
M_{12}^{LQ} &= & \frac{{(\lambda^{32 }}^{*} {\lambda^{2 2}})^2}{192 \pi^2 M_S^2} \eta_B \hat B_{B_s} f_{B_s}^2 M_{B_s}\;,~~~~~~{\rm for}~X(3,2,1/6)\nn\\
M_{12}^{LQ} &= & \frac{(\lambda^{32} {\lambda^{2 2}}^*)^2}{384 \pi^2 M_S^2} \eta_B \hat B_{B_s} f_{B_s}^2 M_{B_s}\;,~~~~~~{\rm for}~X(3,2,7/6).
\eea
Including both the SM and leptoquark contributions the total  mass difference is 
given as
\be
\Delta M_s = \Delta M_s^{SM} \left | \left [1 + \frac{c}{16 G_F^2 |V_{tb} V_{ts}^*|^2 m_W^2 S_0(x_t)}
 \left (\frac{(\lambda^{32} {\lambda^{22}}^{*})^2}{M_S^2}  \right )\right ]\right |\;,
\ee 
where the constant $c= 1 $ for $X(3,2,1/6)$ and $1/2$ for  $X(3,2,7/6)$.  Now varying the  mass difference $(\Delta M_s/\Delta M_s^{SM}) $
   within its 
$1 \sigma $ allowed range \cite{pdg},   the constraint on $|\lambda^{32} \lambda^{22}/M_S|$ is  found to be  \cite{mohanta-Bs-mixing}
\bea
0 \leq \left |\frac{\lambda^{32} \lambda^{22}}{M_S} \right |\leq 7.5 \times 10^{-5}~{\rm GeV}^{-1}\;,~~~~~~~~ {\rm for}~X(3,2,7/6),\nn\\
0 \leq \left |\frac{\lambda^{32} \lambda^{22}}{M_S} \right | \leq 5.0 \times 10^{-5}~{\rm GeV}^{-1}\;,~~~~~~~~~ {\rm for}~X(3,2,1/6).\label{scale}
\eea 
In order to relate this results with the bounds obtained $B_s \to \mu \mu $ process, we  scale the couplings obtained from $B_s - \bar B_s$ mass difference  for a benchmark  leptoquark mass of 1 TeV and the bounds in Eq. 
(\ref{scale})  is translated  as 
\bea
0 \leq \left |\frac{\lambda^{32} \lambda^{22}}{M_S^2}\right | 
 \leq 7.5 \times 10^{-8}~{\rm GeV}^{-2},~~~~~~~~ {\rm for}~X(3,2,7/6)\nn\\
0 \leq \left |\frac{\lambda^{32} \lambda^{22}}{M_S^2}\right | 
\leq 5.0 \times 10^{-8}~{\rm GeV}^{-2},~~~~~~~~ {\rm for}~X(3,2,1/6),
 \eea
which are reasonably higher than those of  obtained from $B_s \to \mu \mu$ process. Hence in our analysis,
we will use the bounds (\ref{r-bound1}) as discussed in the previous subsection.
%%%%%%%%%%%%%%%%%%%%%%%%%%%%%%%%%%%%%%%%%%%%%%%%%%%%%%%%%%%%%%%%%%%%%%%%%%%%%%%
\section{Numerical Analysis}
After having the detailed knowledge about the SM observables and the bound on the new leptoquark couplings, we now proceed for numerical analysis. We have taken  the particle masses and the life time of $\Lambda_b$ baryon from \cite{pdg}.  
%%%%%%%%%%%%%%%%%%%%%%%%%%%%%%%%%%%%%%%%%%%%%%%%%%%%%%%%%%%%%%%%%%%%%%%%%
\begin{table}[h]
\caption{Numerical values of the form factor $f_1(0)$, $f_2(0)$ and the parameters involved in the double fit for $\Lambda_b \to \Lambda$ transition.}
\begin{center}
\begin{tabular}{|c|  c|  }
 \hline

 Parameter & ~~LCSR (twist-3) \cite{LCSR} ~~ \\

 \hline
 
$f_1(0)$ & $0.14^{+0.02}_{-0.01} $  \\

$a$ & $2.91^{+0.1}_{-0.07}$ \\

$b$ & $2.26^{+0.13}_{-0.08}$  \\

\hline

~~$f_2(0)~(10^{-2} ~{\rm GeV^{-1}})$~~ & $-0.47^{+0.06}_{-0.06} $  \\

$a$ & $3.4^{+0.06}_{-0.05}$ \\

$b$ & $2.98^{+0.09}_{-0.08}$  \\
\hline
\end{tabular}
\end{center}
\end{table}
%%%%%%%%%%%%%%%%%%%%%%%%%%%%%%%%%%%%%%%%%%%%%%%%%%%%%%%%%%%%%%%%%%%%%%%
The $q^2$ dependence of form factors derived in the light cone sum rule (LCSR) approach can be  parameterized as
\bea
f_i (q^2) = \frac{f_i (0)}{1- a (q^2/ m_{\Lambda_b}^2) +  b (q^2/ m_{\Lambda_b}^2)^2},
\eea
where the values of the parameters $f_i(0)$, $a$  and $b$ and are listed in Table II \cite{LCSR}. The other form factors are related to these two and the HQET form factors $(F_{1, 2})$ through \cite{LCSR}
\bea
&&f_2^T = g_2 ^T =f_1 = g_1 = F_1 + \frac{m_\Lambda}{m_{\Lambda_b}} F_2, \nn \\
&&f_2 = g_2 = f_3 =  g_3 = \frac{F_2}{m_{\Lambda_b}}, \nn \\
&&f_1^T = g_1^T = \frac{F_2}{m_{\Lambda_b}} q^2. 
\eea
In the lattice QCD formalism, the $\Lambda_b \to \Lambda$ helicity  form factors, i.e.,  $f_{+, \perp, 0}$, $g_{+, \perp, 0}$, $h_{+, \perp}$ and $\tilde{h}_{+, \perp}$ in the physical limit can have the simple form \cite{Latt-QCD-2} 
\bea
f(q^2) = \frac{1}{1-q^2/(m_{pole}^f)^2} \Big [ a_0^f +a_1^f z(q^2) + a_2^f [z(q^2)]^2 \Big ],
\eea
where the values and uncertainties of the parameters $a_0^f$, $a_1^f$ and $a_2^f$ from the higher-order fit are given in Table V of \cite{Latt-QCD-2}. These helicity form factors are related to the form factors $f_i^{(T)}$ and $g_i^{(T)}$ used in this work as follows:
\begin{eqnarray}
f_{+}&=&f_{1}-\frac{q^{2}}{m_{\Lambda_{b}}+m_{\Lambda}}f_{2},~~~~~ f_{\bot}=f_{1}-(m_{\Lambda_{b}}+m_{\Lambda})f_{2},~~~~~ f_{0}=f_{1}+\frac{q^{2}}{m_{\Lambda_{b}}-m_{\Lambda}}f_{3},\nonumber \\g_{+}&=&g_{1}+\frac{q^{2}}{m_{\Lambda_{b}}-m_{\Lambda}}g_{2},~~~~~
g_{\bot}=g_{1}+(m_{\Lambda_{b}}-m_{\Lambda})g_{2}, ~~~~~
g_{0}=g_{1}-\frac{q^{2}}{m_{\Lambda_{b}}+m_{\Lambda}}g_{3},\nonumber \\h_{+}&=&f_{2}^{T}-\frac{m_{\Lambda_{b}}+m_{\Lambda}}{q^{2}}f_{1}^{T},~~~~~h_{\bot}=f_{2}^{T}-\frac{f_{1}^{T}}{m_{\Lambda_{b}}+m_{\Lambda}},\nonumber \\
\tilde{h}_{+}&=&g_{2}^{T}+\frac{m_{\Lambda_{b}}-m_{\Lambda}}{q^{2}}g_{1}^{T},~~~~~\tilde{h}_{\bot}=g_{2}^{T}+\frac{g_{1}^{T}}{m_{\Lambda_{b}}-m_{\Lambda}}.
\end{eqnarray}
In our analysis, we have taken the form factors computed in the light cone sum rule approach for low $q^2$ region (as these are not so well-behaved in the high $q^2$ regime), and for high $q^2$ theory we have used the lattice QCD calculations of $\Lambda_b \to \Lambda$  form factors \cite{Latt-QCD-2}. 
The values of the Wilson coefficients used in our analysis are evaluated at the renormalization scale $\mu \approx m_b = 4.8$ GeV. 
 In the LQ model, the  new physics contributions to the branching ratios and forward-backward asymmetry parameters are encoded in the new Wilson coefficients. By using the above input parameters and the values of the new Wilson coefficients, we show in Fig. 1, the $q^2$ variation of branching ratio of  $\Lambda_b \to \Lambda e^+ e^-$ (top left panel), $\Lambda_b \to \Lambda \mu^+ \mu^-$ (top right panel) and $\Lambda_b \to \Lambda \tau^+ \tau^-$ (bottom panel) processes  for the full kinematically  accessible  physical region.   In these plots, we have shown the contributions arising from the exchange  of $X=(3, 2, 7/6)$ leptoquark. The  SM contributions are represented by blue lines and the grey bands denote the theoretical  uncertainties arising due to the uncertainties associated with the CKM matrix elements and the hadronic form factors. The green bands represent the leptoquark contributions to the branching ratios. The bin-wise experimental results for
$\Lambda_b \to \Lambda \mu^+ \mu^-$ process \cite{lhcb-lambda} are shown by black data points. There is slight deviation in the decay distribution between the predicted and observed data. The corresponding results  coming from the exchange of the  $X=(3, 2, 1/6)$ LQ are shown in  Fig. 2. 
From these figures, one can  see that the branching ratios of 
$\Lambda_b \to \Lambda e^+ e^-$  and $\Lambda_b \to \Lambda \tau^+ \tau^-$ decay processes deviate significantly  from their SM predictions, whereas the new physics effects on $\Lambda_b \to \Lambda \mu^+ \mu^-$  branching ratio is  not so prominent. 
In Table III, we present the integrated values of branching ratio for all the above processes, where we have used the veto windows as $(8~{\rm GeV}^2 <m_{l^+ l^-}^2<11~{\rm GeV}^2)$ and $(12.5~{\rm GeV}^2 <m_{l^+ l^-}^2<15~{\rm GeV}^2)$ \cite{lhcb-lambda}, to eliminate the backgrounds coming from the dominant resonances $\Lambda_b \to \Lambda J/\psi(\psi')$ with
$J/\psi(\psi')\to l^+ l^-$. The predicted branching ratio for $\Lambda_b
\to \Lambda \mu^+ \mu^-$ is  almost consistent with the observed data
 ${\rm Br} (\Lambda_b \to \Lambda \mu^+ \mu^-)=(1.08 \pm 0.28)
 \times 10^{-6}$ \cite{pdg}. Also, as seen from Table III, 
 the experimental result can be accommodated in the  leptoquark model.
 Within the SM, the forward backward asymmetry parameters in the $B \to K l^+ l^-$ decay processes are identically zero since they only involve scalar and tensor types of currents, whereas  $B \to K l^+ l^-$ processes are described by only vector-type interactions.  However, for semileptonic  $\Lambda_b \to \Lambda l^+ l^-$ decay processes, the forward backward asymmetry depends  on two combinations of the Wilson coefficients ${\rm Re}(C_7^{eff} C_{10}^*)$ and ${\rm Re}(C_9^{eff} C_{10}^*)$ \cite{chen1} and thus, can have negative values in the SM. The contribution due to the  new  Wilson coefficients ($C_{9, 10}^{NP(\prime)}$) may enhance the rate of asymmetries and can shift the zero position of these asymmetries.
In Fig. 3, the variation of forward-backward asymmetry  for $\Lambda_b \to \Lambda \mu^+ \mu^-$ (left panel), $\Lambda_b \to \Lambda \tau^+ \tau^-$ (right panel) modes are depicted with respect to  $q^2$ both in the SM and in the $X=(3, 2, 7/6)$ LQ model  including the LD contributions and the corresponding integrated values are presented in Table III. Similarly the variation of forward-backward asymmetries  for $X=(3, 2, 1/6)$ LQ exchange are 
shown  in Fig.  4.  We found no significant deviation of the zero position of $A_{FB}$  from its SM value due to the leptoquark contributions in $\Lambda_b \to \Lambda \mu^+ \mu^-$ process. However, there is certain discrepancy between the observed and  predicted results  in the  high $q^2$ regime.  The forward-backward asymmetry for $\Lambda_b \to \Lambda \tau^+ \tau^-$ process however,  has significant deviation from the SM in both the $X=(3, 2, 7/6)$ and $X=(3, 2, 1/6)$ leptoquark model.
%%%%%%%%%%%%%%%%%%%%%%%%%%%%%%%%%%%%%%%%%%%%%%%%%%%%%%%%%%%%%%%%%%%%%%%%%%%%%%
\begin{figure}[h]
\centering
\includegraphics[scale=0.45]{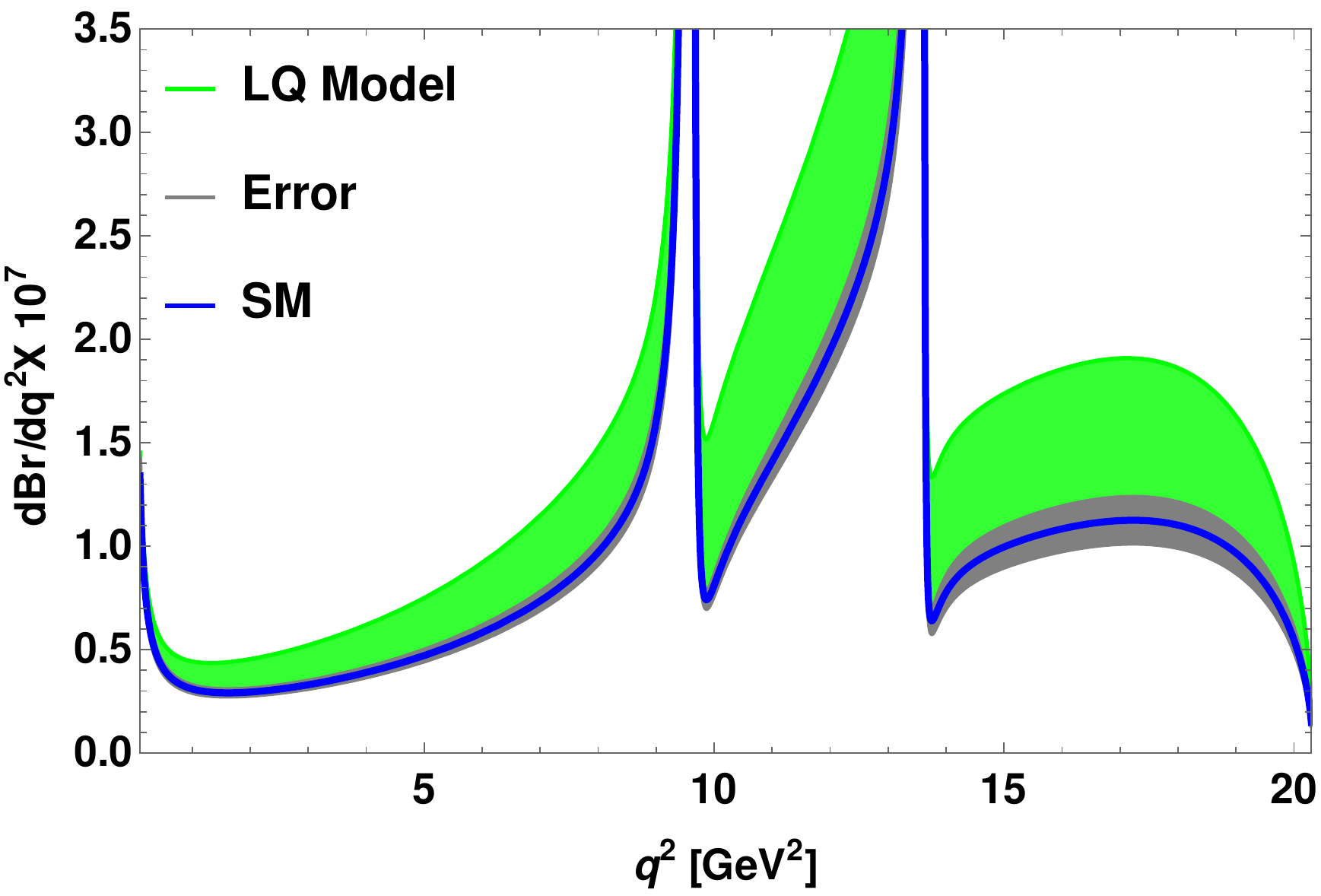}
\quad
\includegraphics[scale=0.45]{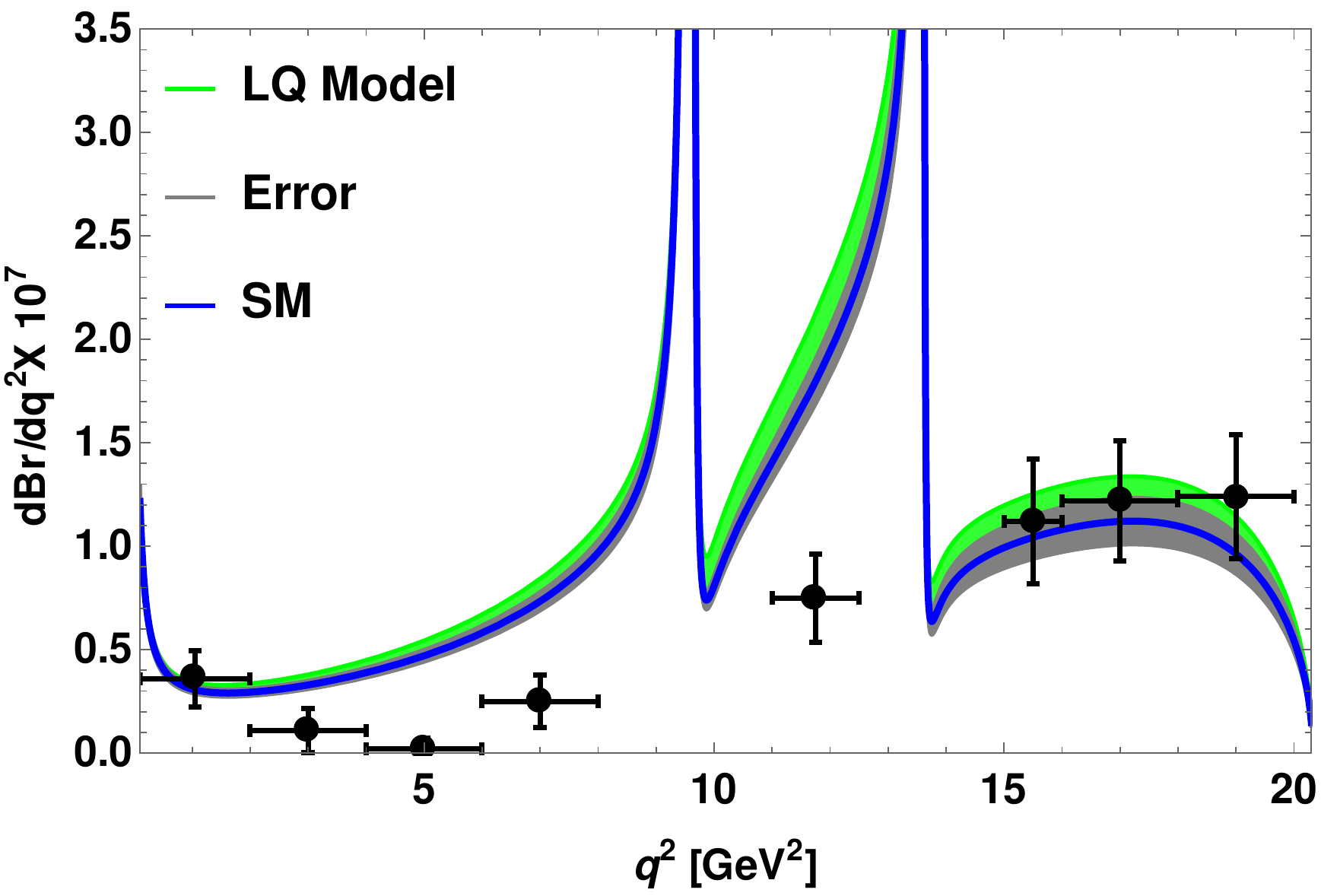}
\quad
\includegraphics[scale=0.45]{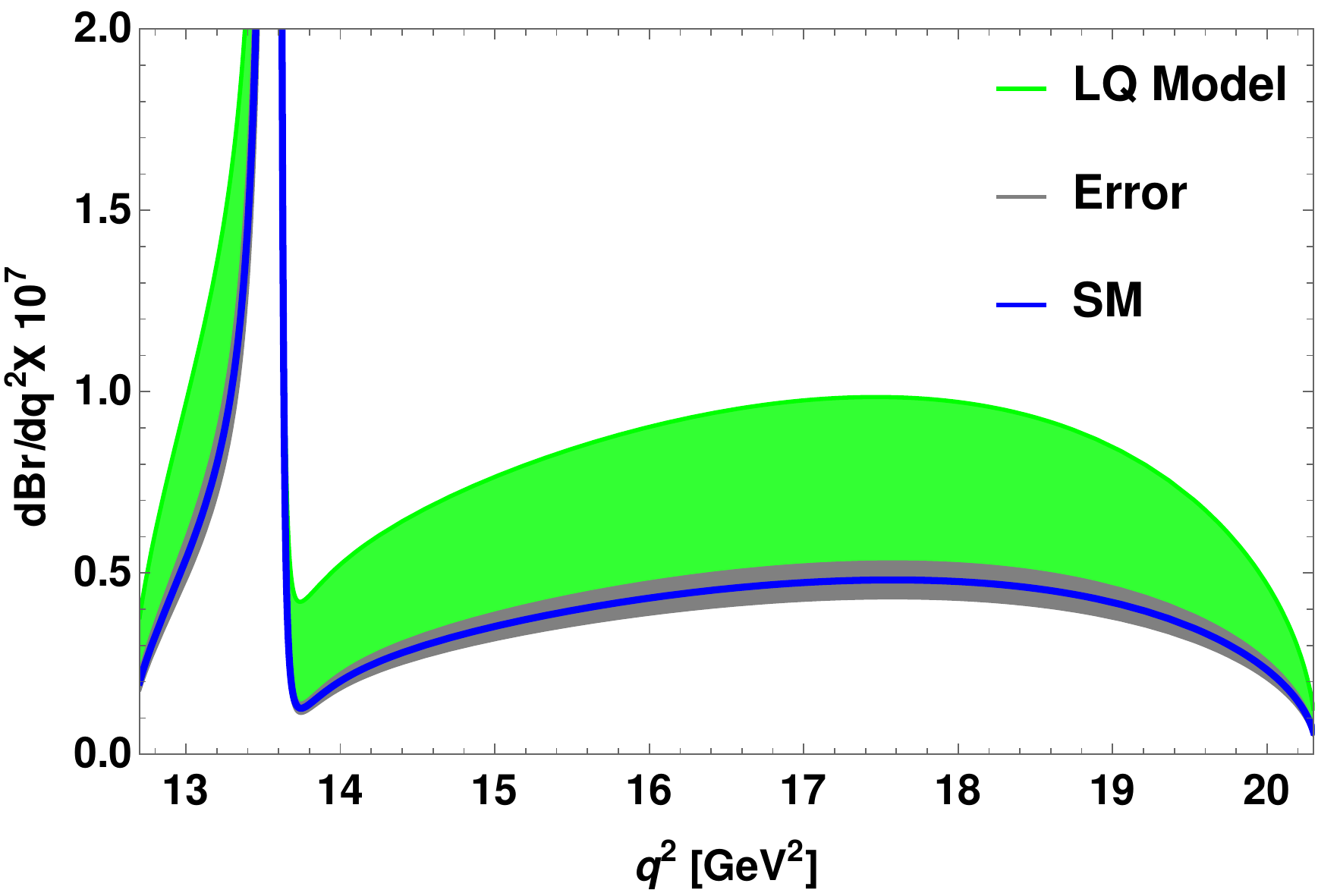}
\caption{The variation of branching ratio of $\Lambda_b \rightarrow \Lambda e^+ e^-$ (left panel), $\Lambda_b \rightarrow \Lambda \mu^+ \mu^-$ (right panel) and  $\Lambda_b \rightarrow \Lambda \tau^+ \tau^-$ (bottom panel)  with respect to low and high $q^2$ including the LD contributions,  both in the SM and in the $X=(3, 2, 7/6)$ leptoquark model. In each plot, the green band represents the leptoquark contribution and the blue solid line is for the SM. The grey band represents the  theoretical uncertainty arises due to the input parameters in the SM. The black data points in $\Lambda_b \to \Lambda \mu^+ \mu^-$ process represent the bin-wise experimental data.}
\end{figure}
%%%%%%%%%%%%%%%%%%%%%%%%%%%%%%%%%%%%%%%%%%%%%%%%%%%%%%%%%%%%%%%%%%%%%%%%%%%%%%
\begin{figure}[h]
\centering
\includegraphics[scale=0.45]{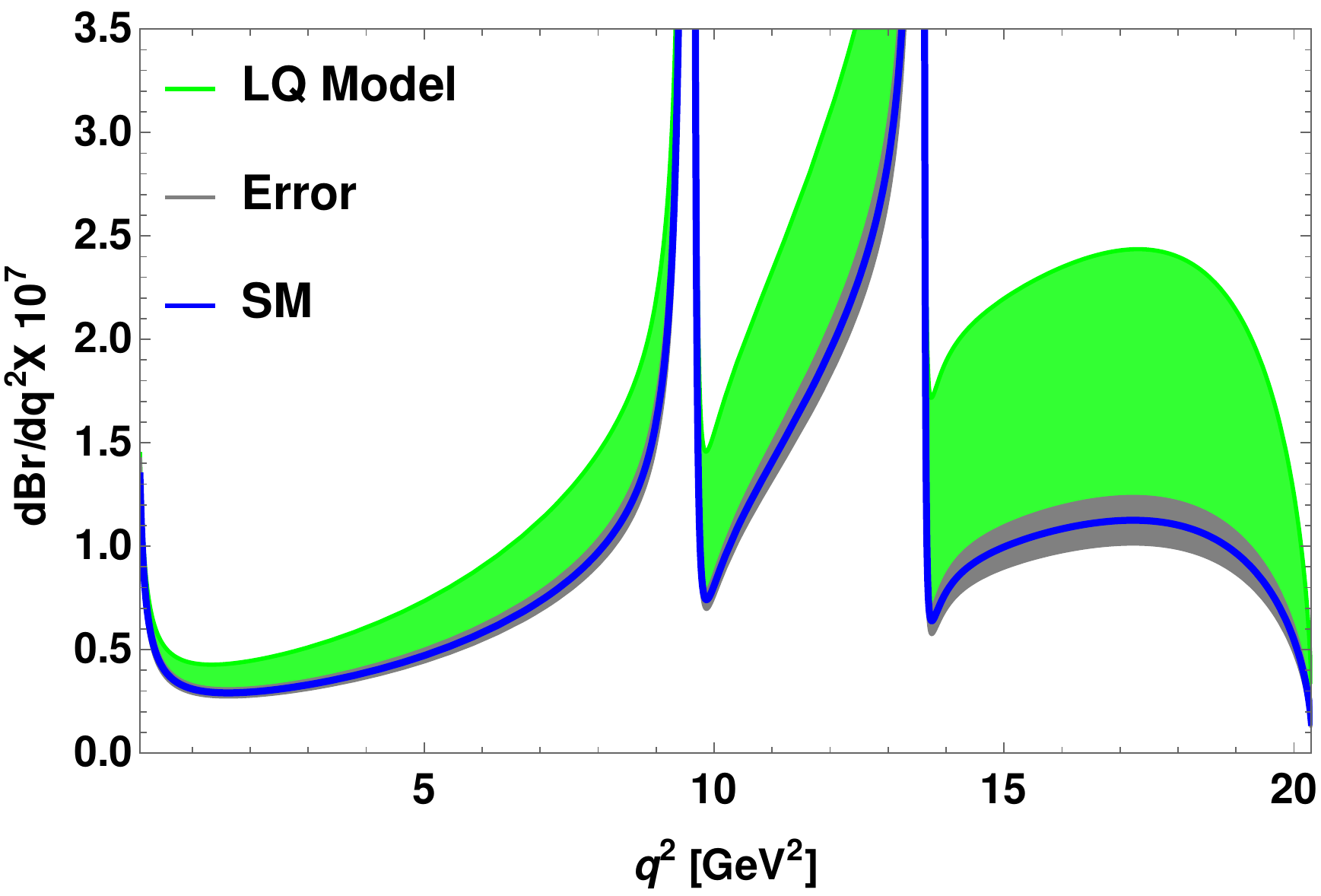}
\quad
\includegraphics[scale=0.45]{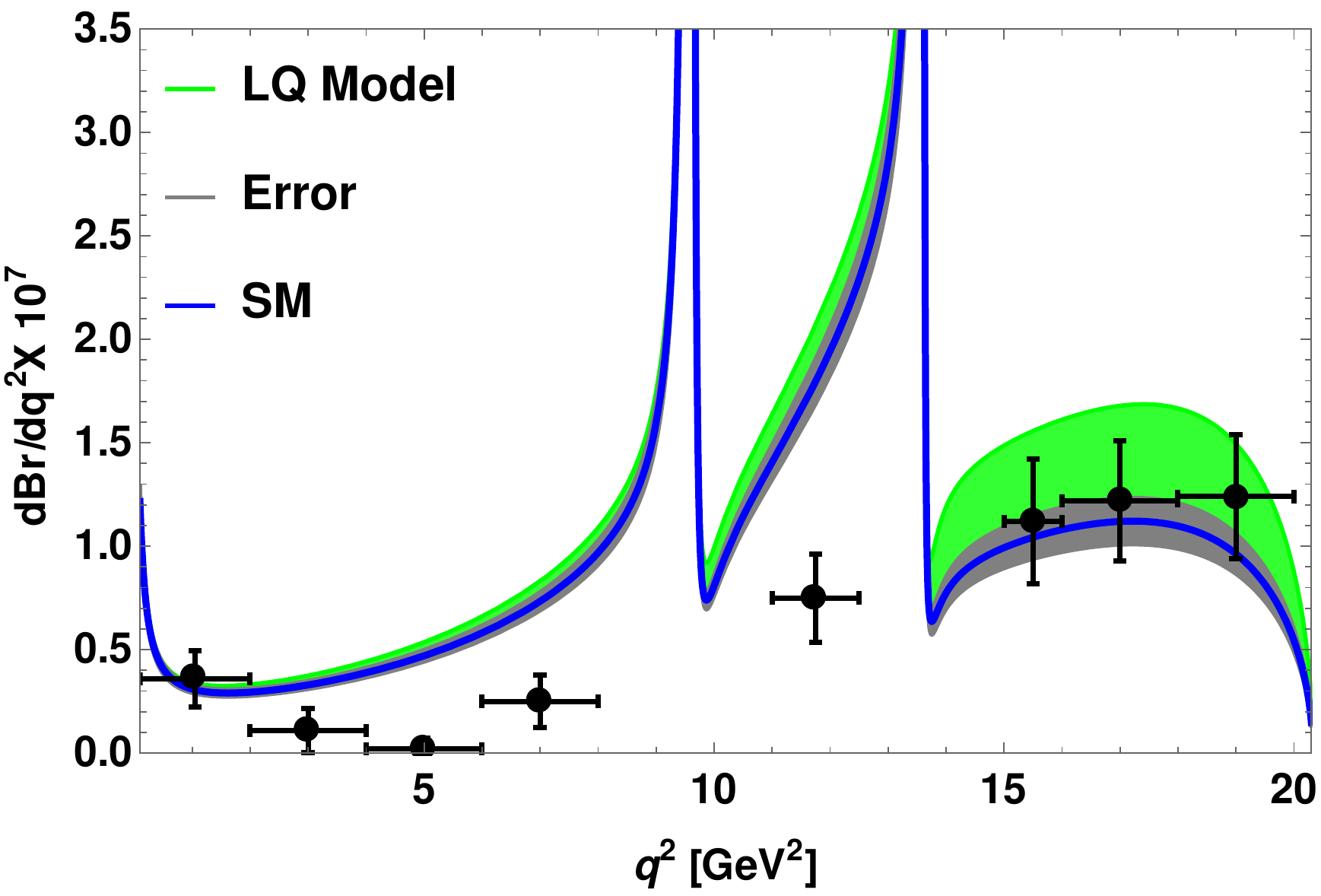}
\quad
\includegraphics[scale=0.45]{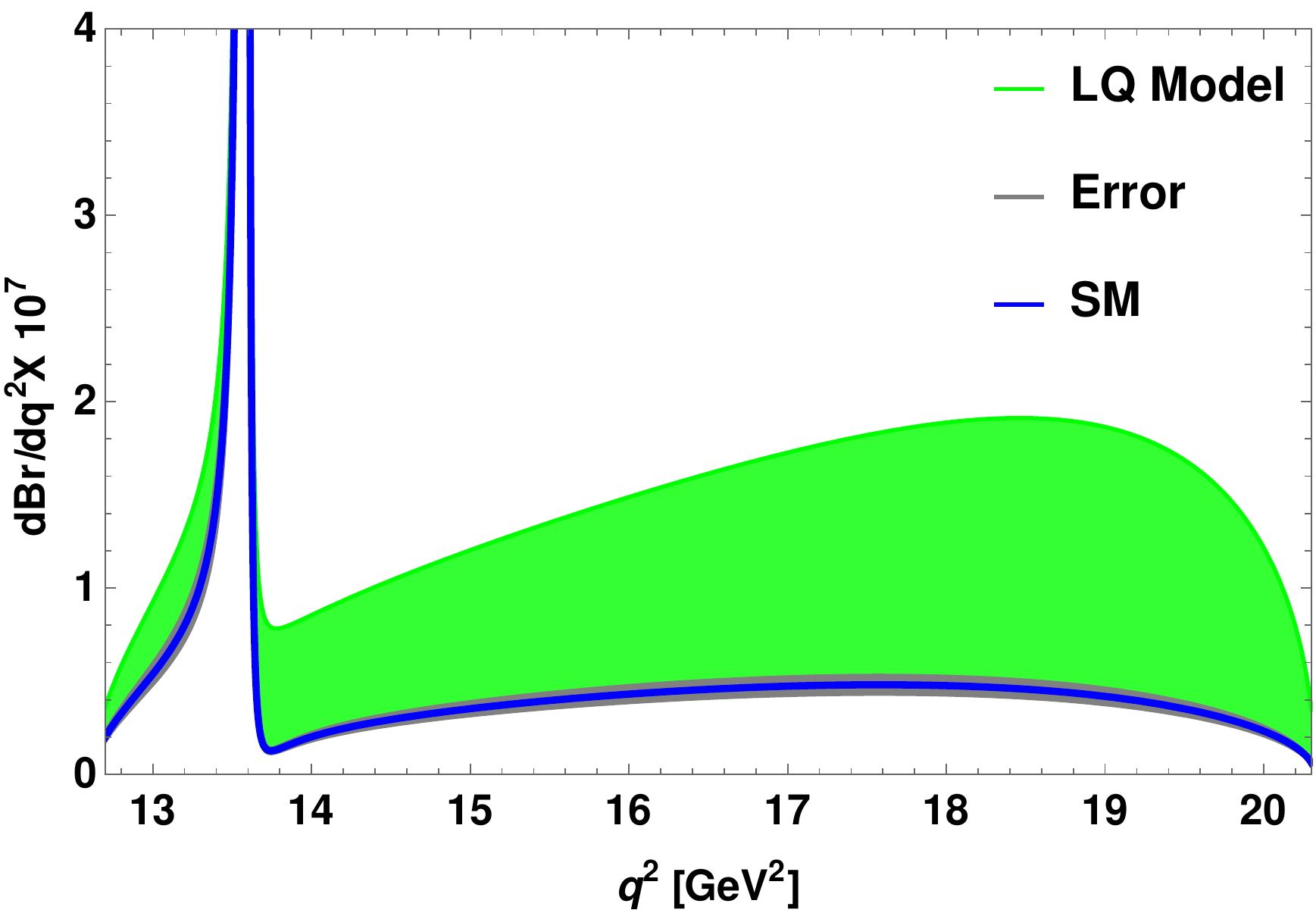}
\caption{Same as Fig.1 for $X=(3, 2, 1/6)$ LQ exchange. }
\end{figure}
%%%%%%%%%%%%%%%%%%%%%%%%%%%%%%%%%%%%%%%%%%%%%%%%%%%%%%%%%%
Besides the branching ratios and forward-backward asymmetry parameters of $\Lambda_b \to \Lambda l^+ l^-$ processes, the new physics effects can also be observed in the lepton polarization asymmetries.
In the left panel of Fig. 5, the distribution of the longitudinal (top), transverse (middle)  and normal (bottom) polarization components for $\Lambda_b \to \Lambda \mu^+ \mu^-$ process are shown both in the SM and in the $X=(3, 2, 7/6)$ LQ model, and the corresponding plots for  $\Lambda_b \to \Lambda \tau^+ \tau^-$  process are presented in the right panel. The integrated values of all the three polarizations in the full physical phase space have been presented in Table III. In Fig. 6, we have shown the variation of  the different polarization parameters for $\Lambda_b \to \Lambda \mu^+ \mu^-$ process in the   $X=(3, 2, 1/6)$ leptoquark model. It is found from Table III, that   the transverse  and normal polarization values are very small in the SM and even the leptoquark model does not give any significant deviation.

 Analogous to the lepton flavour non-universality parameter $R_K$, i.e., the ratio of branching fractions of $B \to K \mu^+ \mu^-$ over $B \to K e^+e^-$, we would like to see whether it is possible to observe  lepton non-universality in the $\Lambda_b$ decays.  We have define these parameters as e.g.,
$R_{\Lambda}^{\mu e}= {\rm Br}(\Lambda_b \to \Lambda \mu^+ \mu^-)/{\rm Br}(\Lambda_b \to \Lambda e^+e^-)$.   
In Fig. 7, we show the variation of  lepton nonuniversality parameter $R_{\Lambda}^{\mu e}$ (top-right panel), $R_{\Lambda}^{\tau e}$ (bottom-left panel) and $R_{\Lambda}^{\tau \mu}$ (bottom-right panel) in their respective $q^2$ region. Also, we show the low-$q^2$ behavior of  $R_{\Lambda}^{\mu e}$ (top-left panel),  in the range $1\leq q^2 \leq 6~ {\rm GeV}^2$. These results are  for $X=(3, 2, 7/6)$ leptoquark. Similarly the lepton nonuniversality plot for $X=(3, 2, 1/6)$ leptoquark exchange is shown in Fig. 8. The integrated values of the lepton non-universality parameter in both SM and LQ model are  presented in Table III. We found that there is significant violation of lepton universality in $\Lambda_b$ decays, though there is no experimental evidence so far. The violation of lepton universality is more pronounced for the processes having $\tau$ as final particle. However, as the reconstruction of tau events are extremely difficult, this observable may not be sensitive enough to be observed in near future. As seen from the top-left panel of Figs. 7 and 8, the parameter $R_\Lambda^{\mu e}$ is very promising for the Belle II experiment, as the LHCb, being a hadronic machine works better in muon mode than electron. 

\begin{figure}[h]
\centering
\includegraphics[scale=0.45]{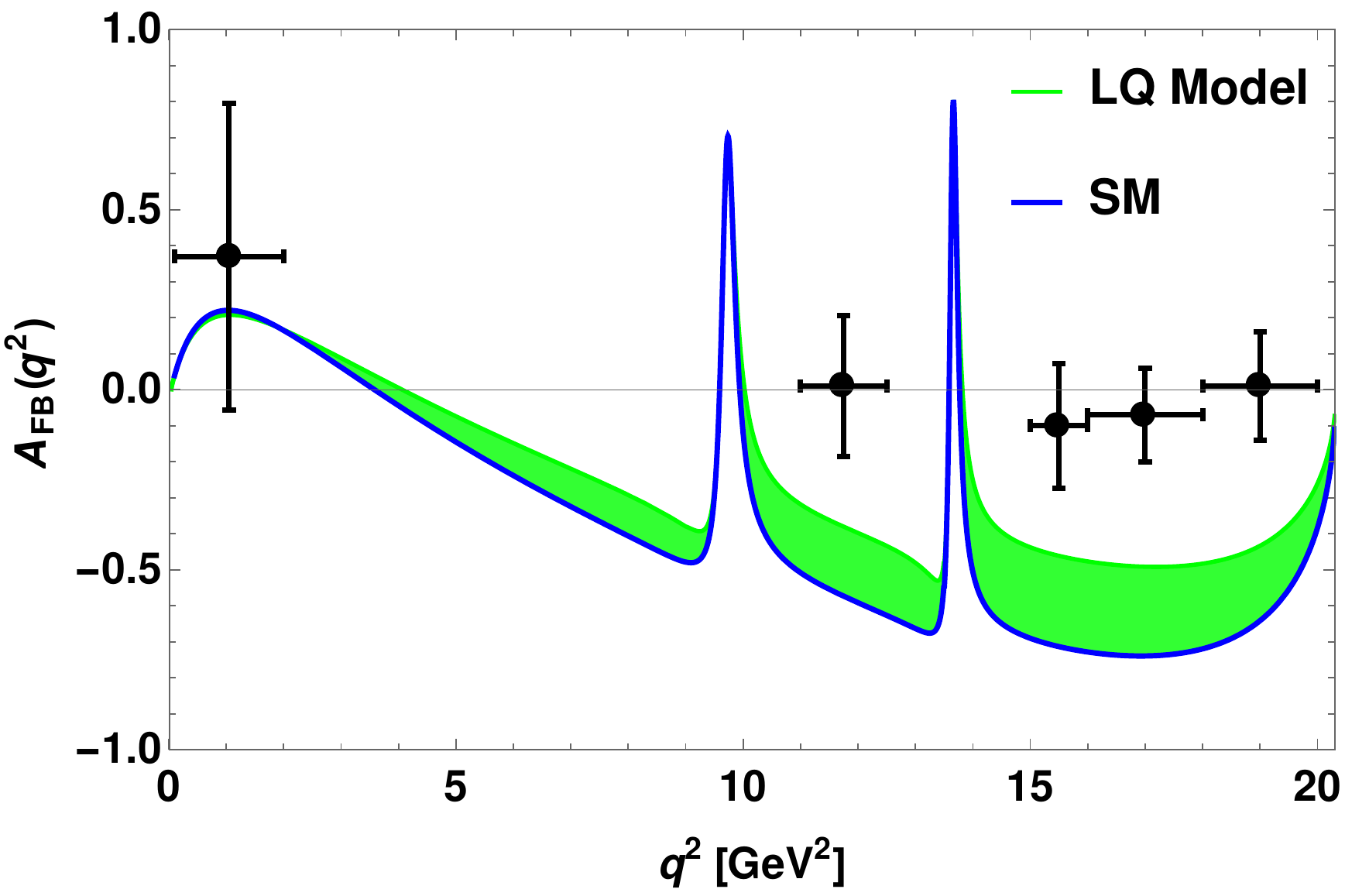}
\quad
\includegraphics[scale=0.45]{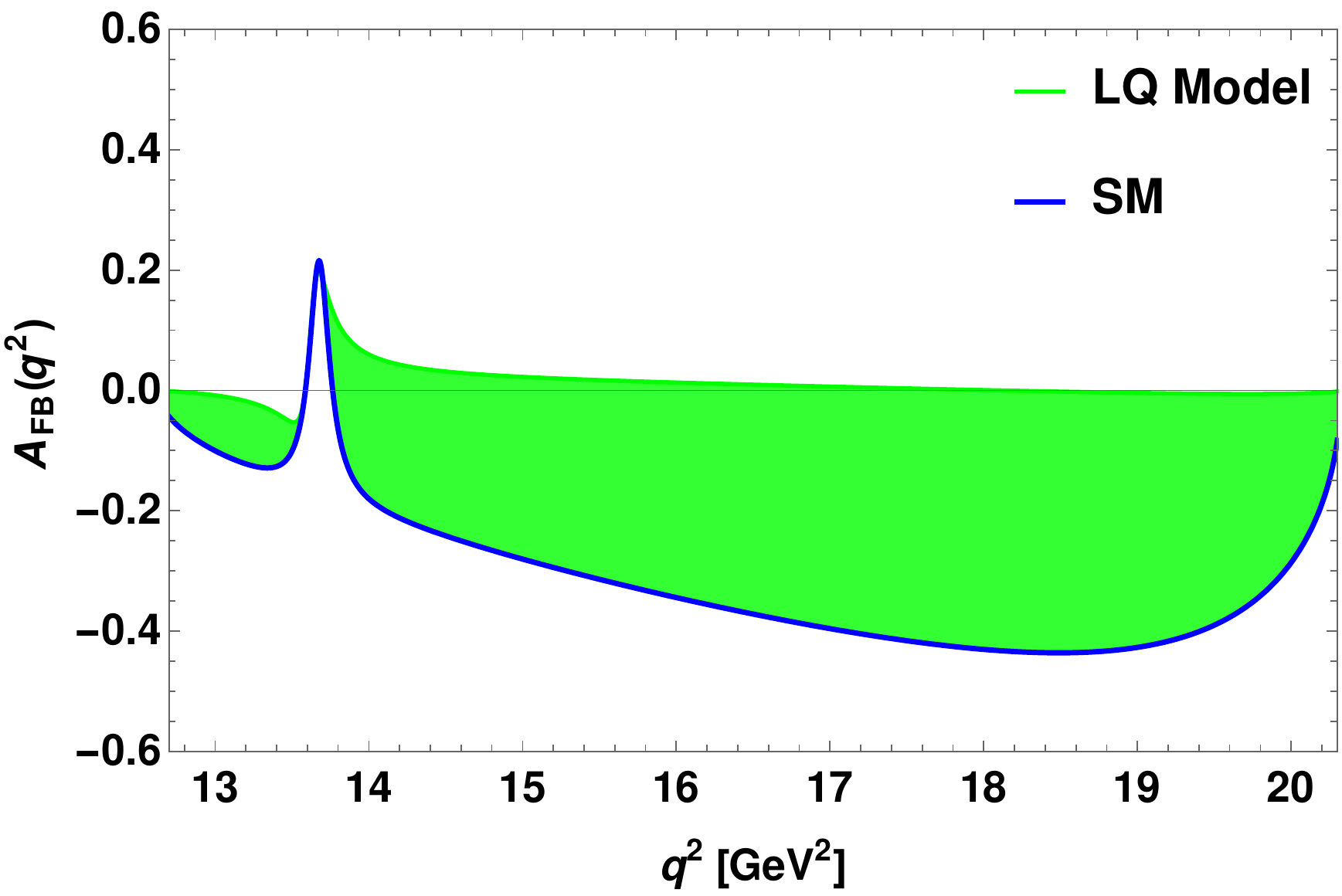}
\caption{The forward-backward asymmetry variation of  $\Lambda_b \rightarrow \Lambda \mu^+ \mu^-$ (left panel) and $\Lambda_b \rightarrow \Lambda \tau^+ \tau^-$ (right panel)  with respect to $q^2$ for $X=(3, 2, 7/6)$ LQ exchange. The black data points in $\Lambda_b \to \Lambda \mu^+ \mu^-$ process represent the bin-wise experimental data. }
\end{figure}
%%%%%%%%%%%%%%%%%%%%%%%%%%%%%%%%%%%%%%%%%%%%%%%%%%%%%%%%%%%%%%%%%%%%%%%%%%%%%%
\begin{figure}[h]
\centering
\includegraphics[scale=0.45]{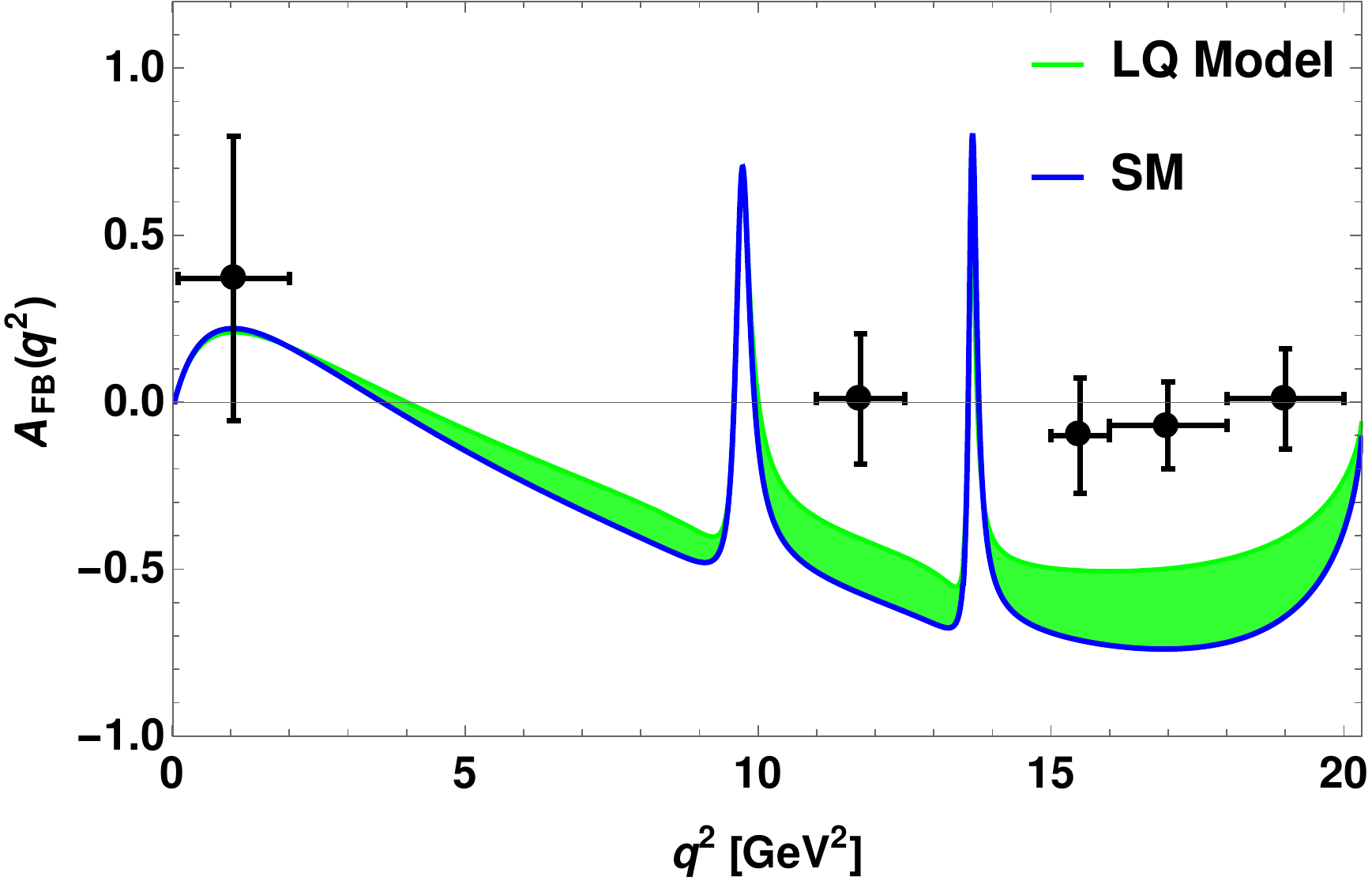}
\quad
\includegraphics[scale=0.45]{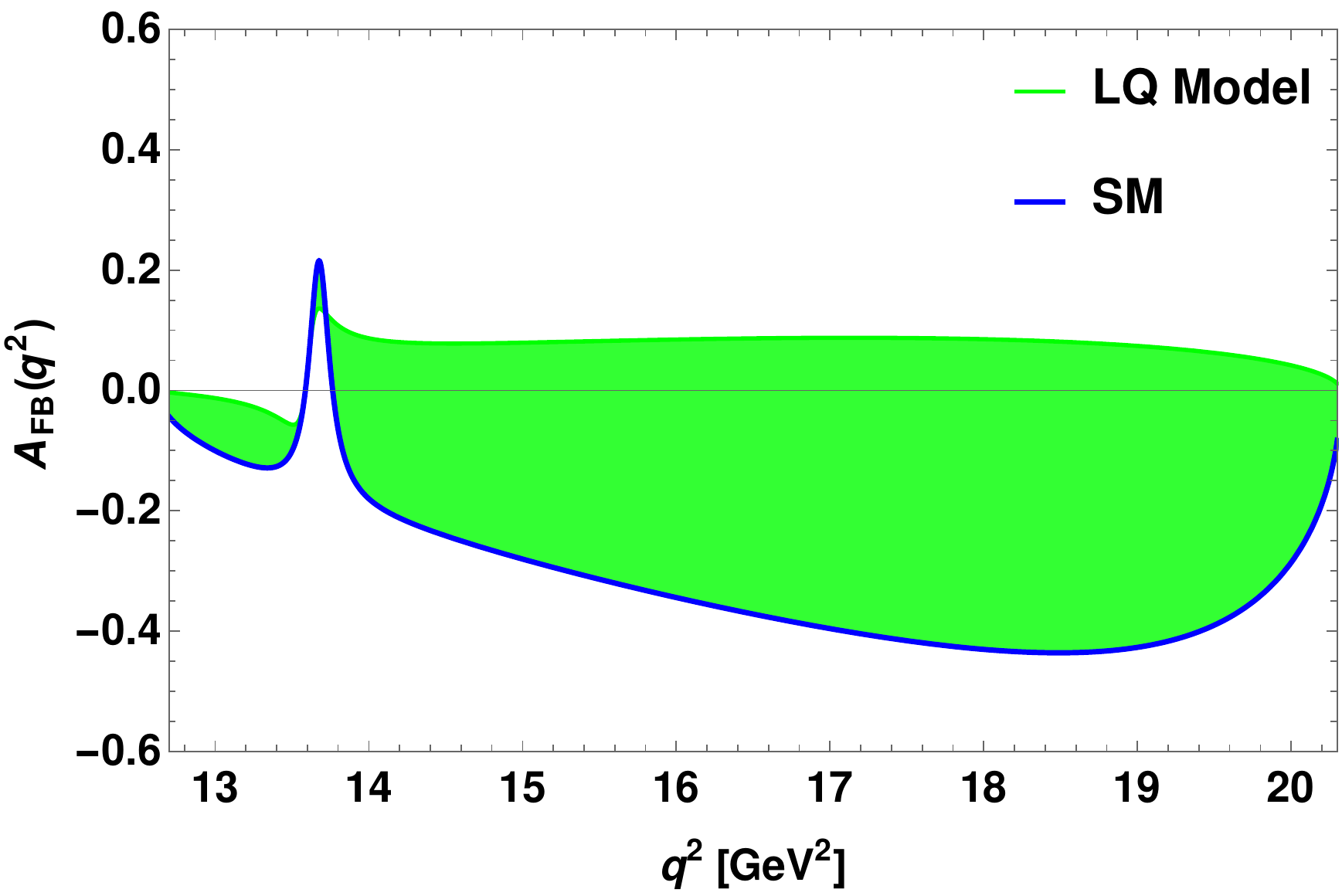}
\caption{Same as Fig.3 for $X=(3, 2, 1/6)$ LQ exchange.}
\end{figure}

%%%%%%%%%%%%%%%%%%%%%%%%%%%%%%%%%%%%%%%%%%%%%%%%%%%%%%%%%%%%%%%%%%%%%%%%%%%%%%
\begin{figure}[h]
\centering
\includegraphics[scale=0.45]{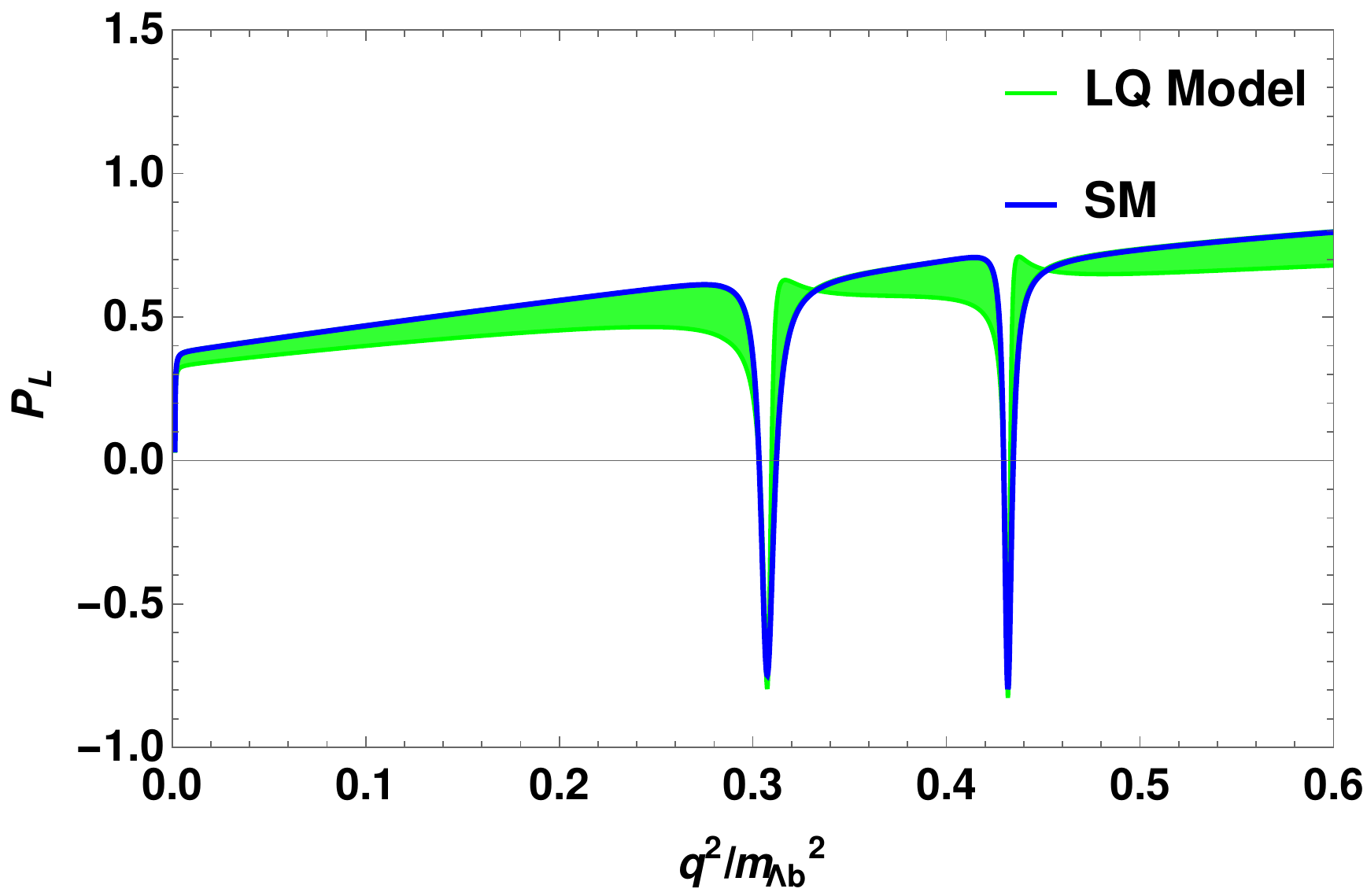}
\quad
\includegraphics[scale=0.45]{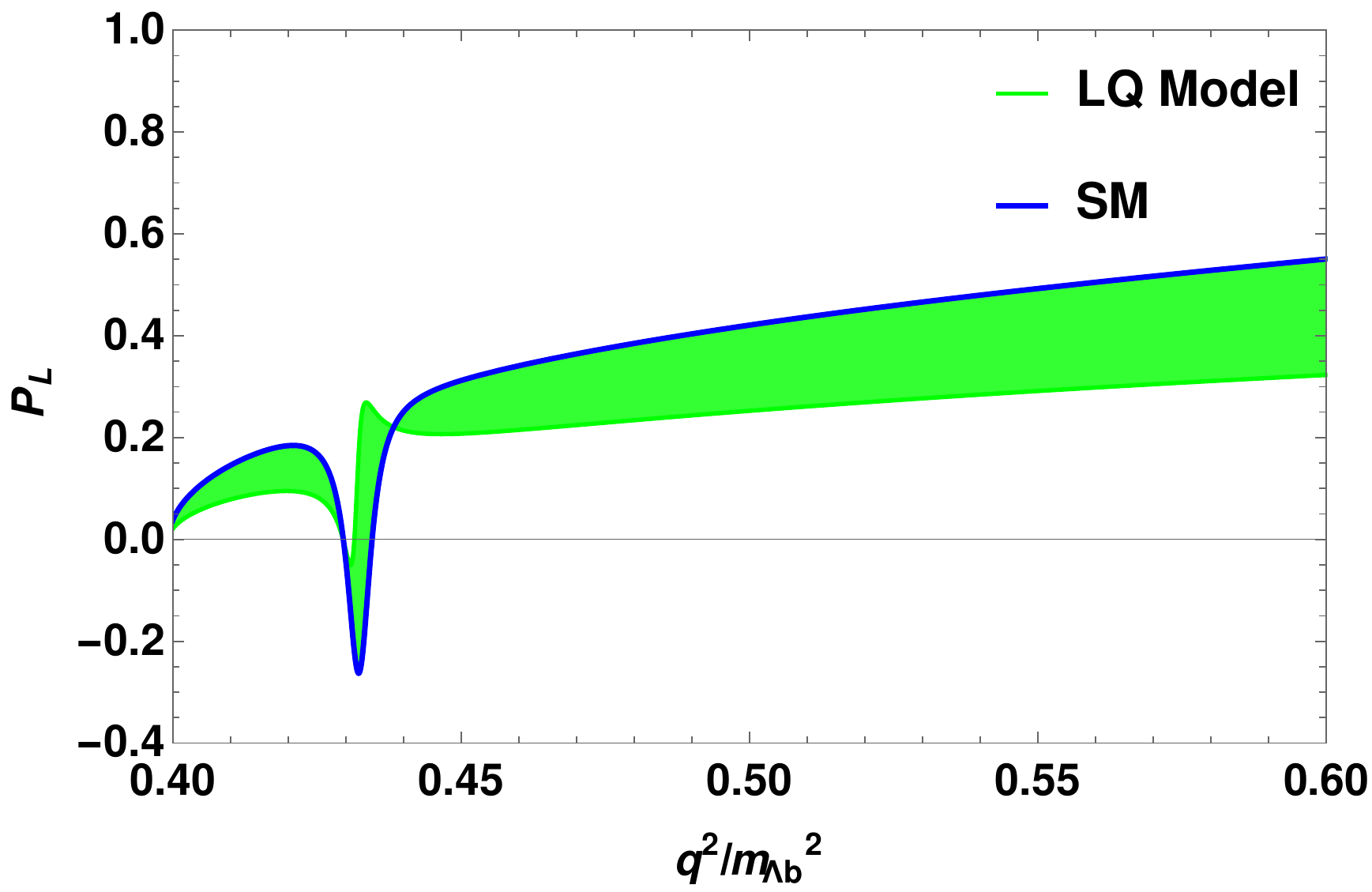}\\
\includegraphics[scale=0.45]{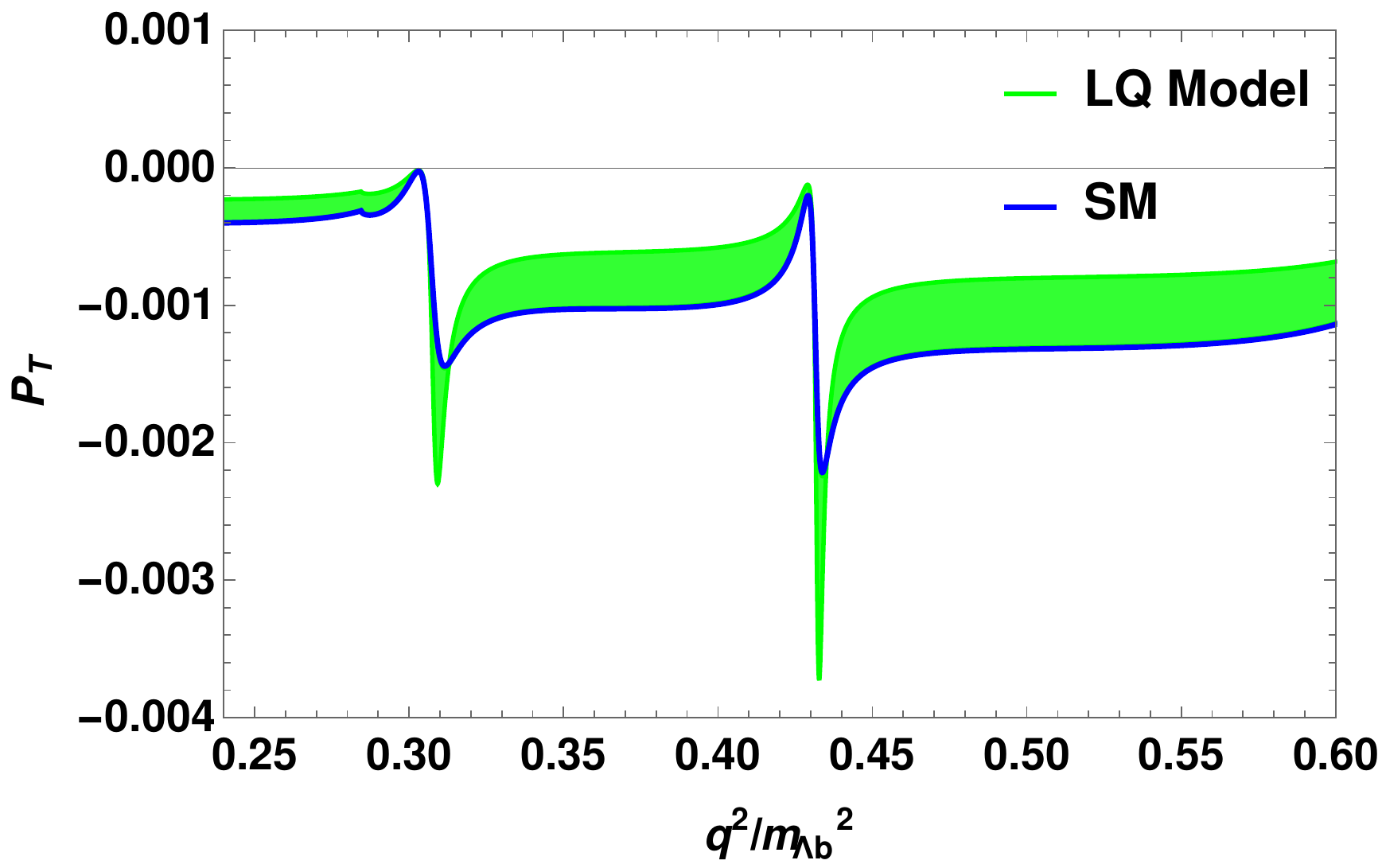}
\quad
\includegraphics[scale=0.45]{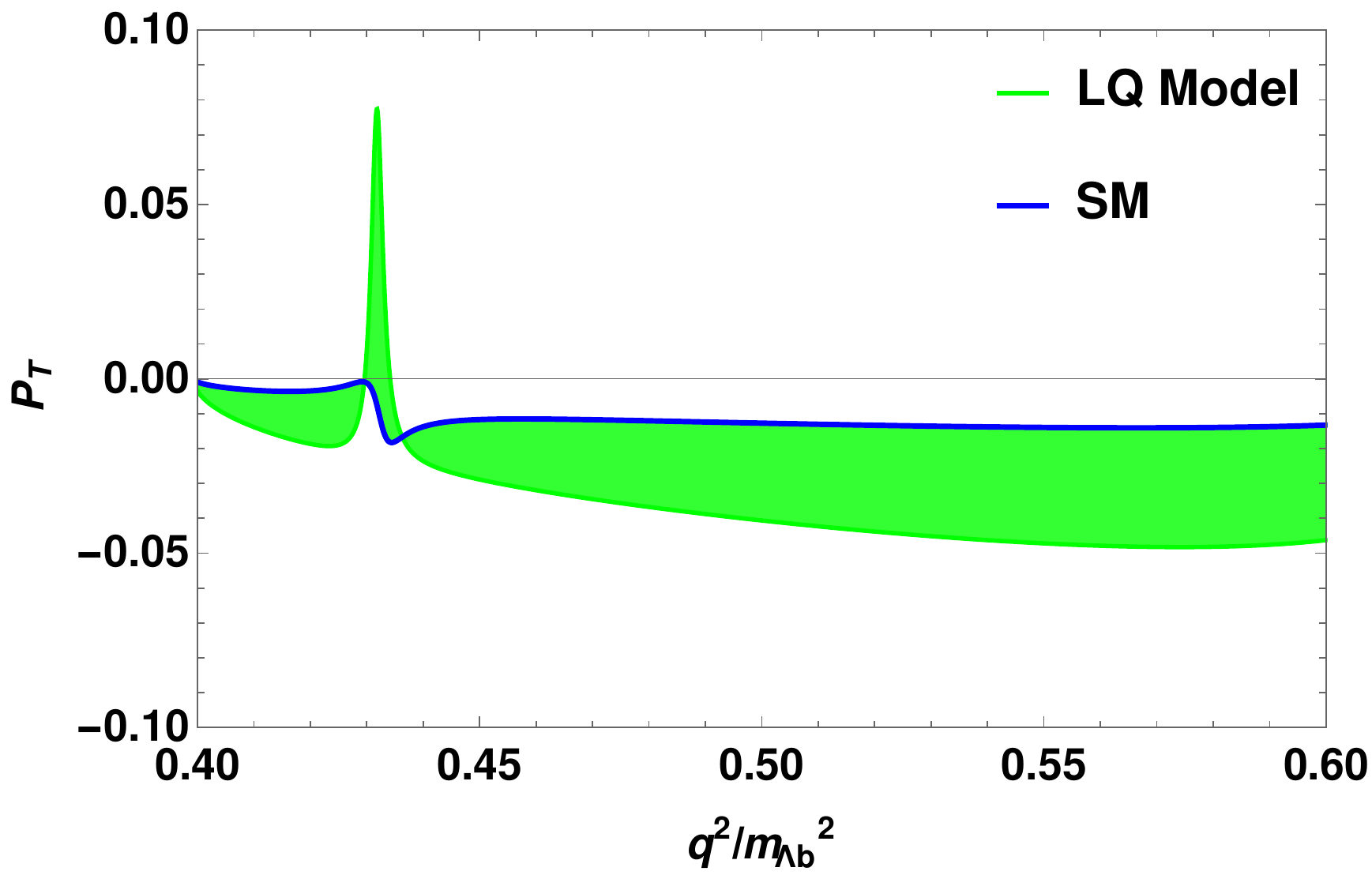}\\
\includegraphics[scale=0.45]{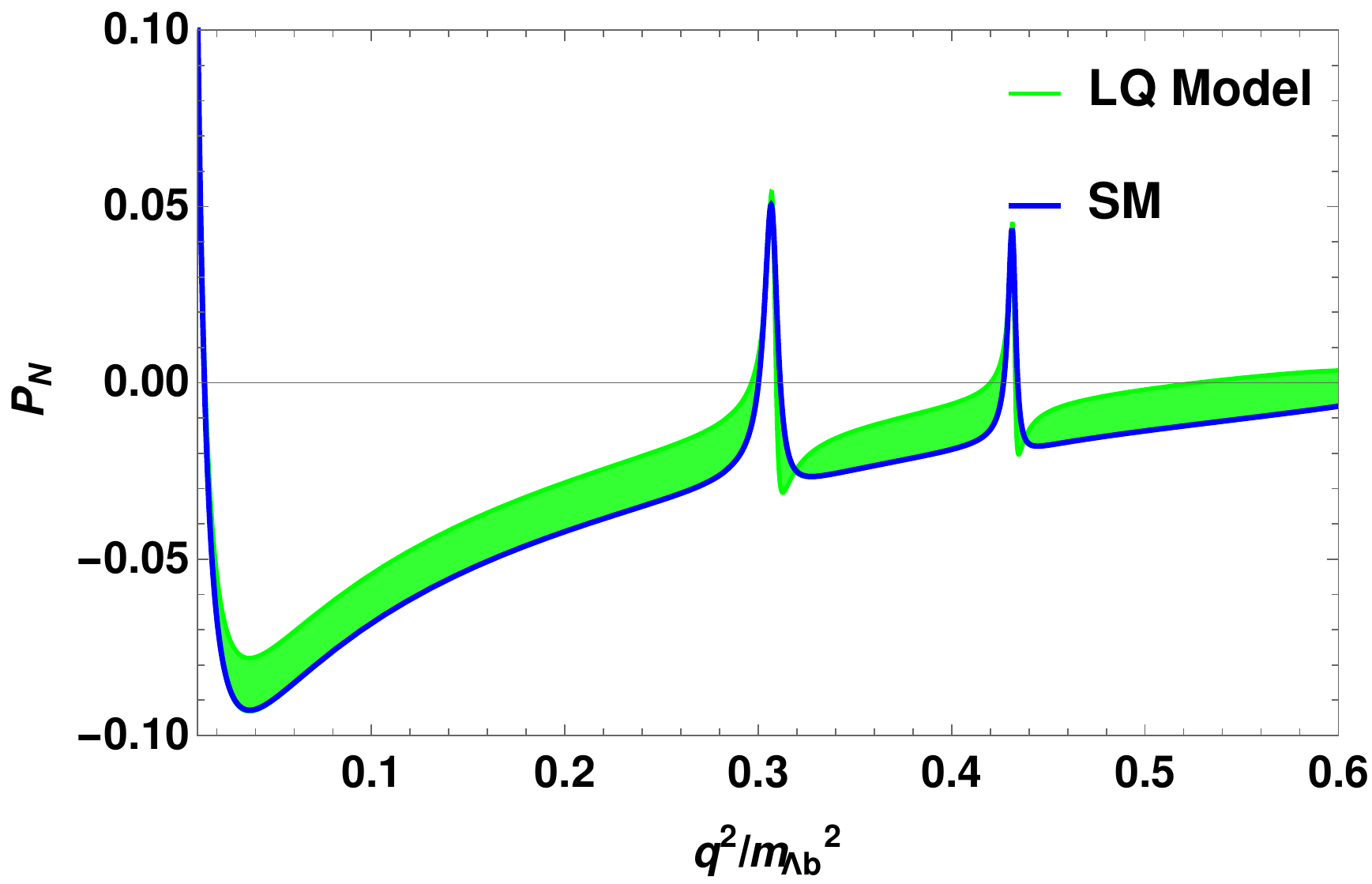}
\quad
\includegraphics[scale=0.45]{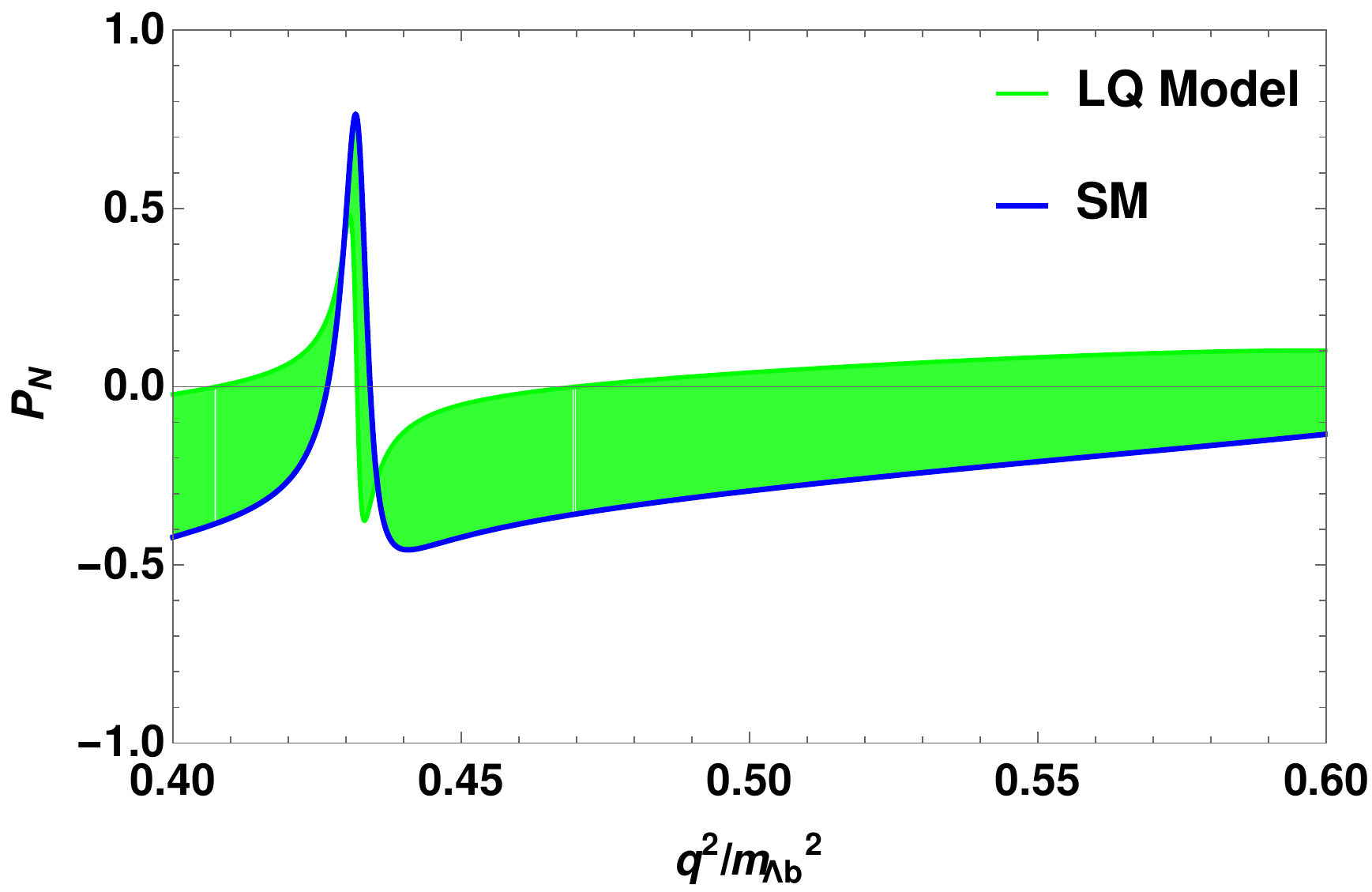}
\caption{The plots in the left panel represent the longitudinal (top), transverse (middle) and normal (bottom) polarizations for $\Lambda_b \rightarrow \Lambda \mu^+ \mu^-$ precess  with respect to $q^2/m_{\Lambda_b}^2$ in the $X=(3, 2, 7/6)$ LQ model. The corresponding   plots for $\Lambda_b \rightarrow \Lambda \tau^+ \tau^-$ mode are shown in the right panel.}
\end{figure}

%%%%%%%%%%%%%%%%%%%%%%%%%%%%%%%%%%%%%%%%%%%%%%%%%%%%%%%%%%%%%%%%%%%%%%%%%%%%%%
\begin{figure}[h]
\centering
\includegraphics[scale=0.45]{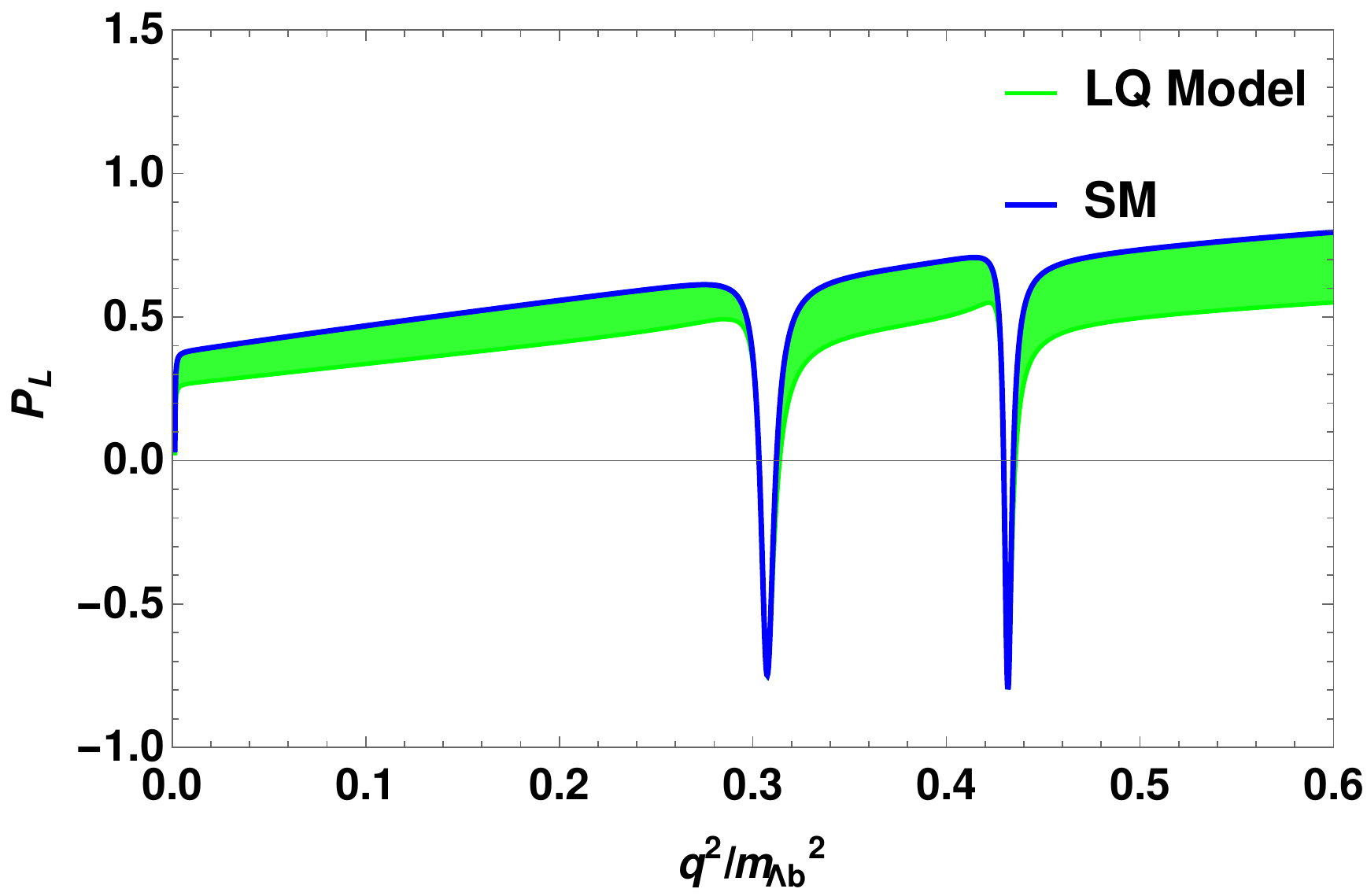}
\quad
\includegraphics[scale=0.45]{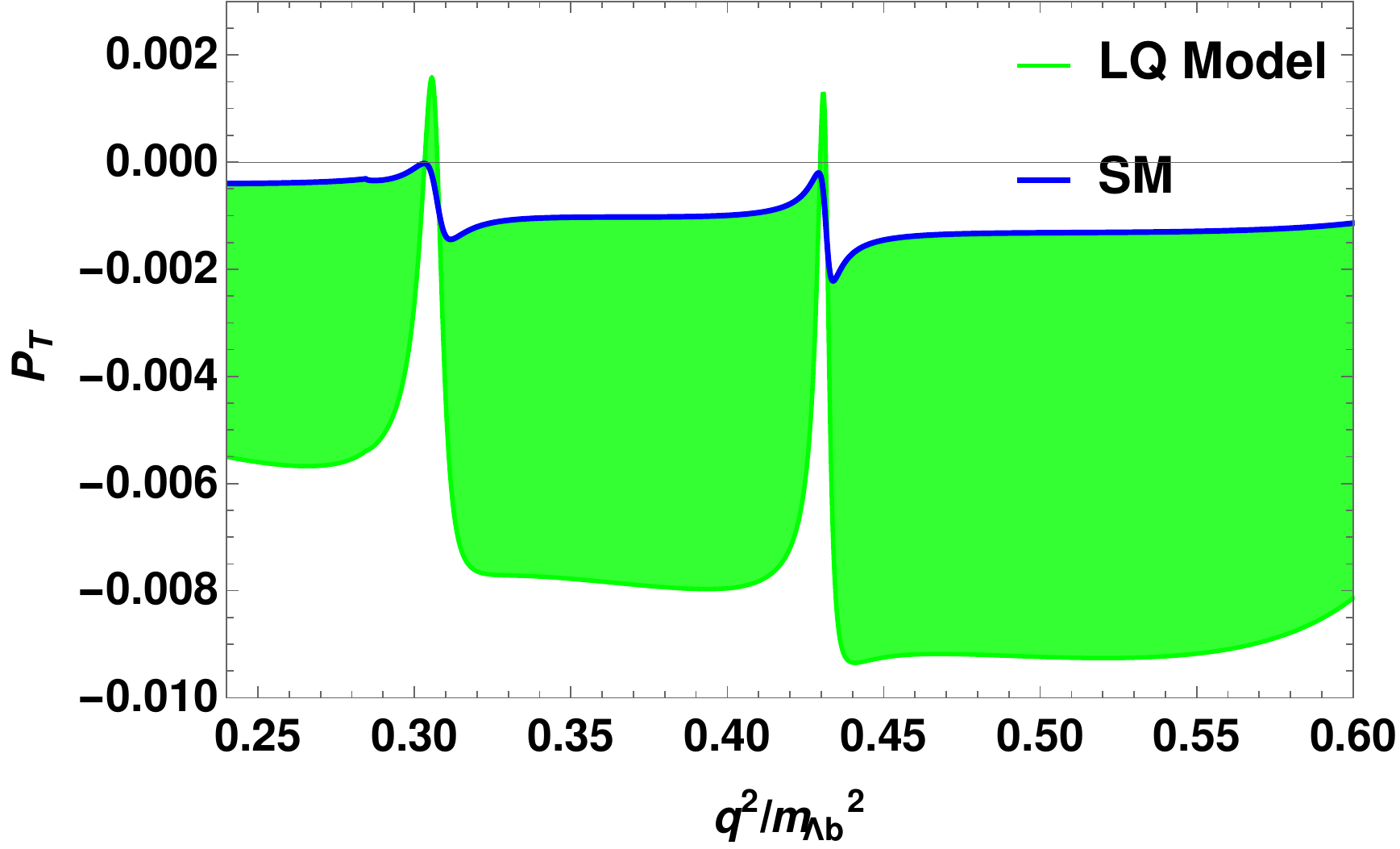}
\quad
\includegraphics[scale=0.45]{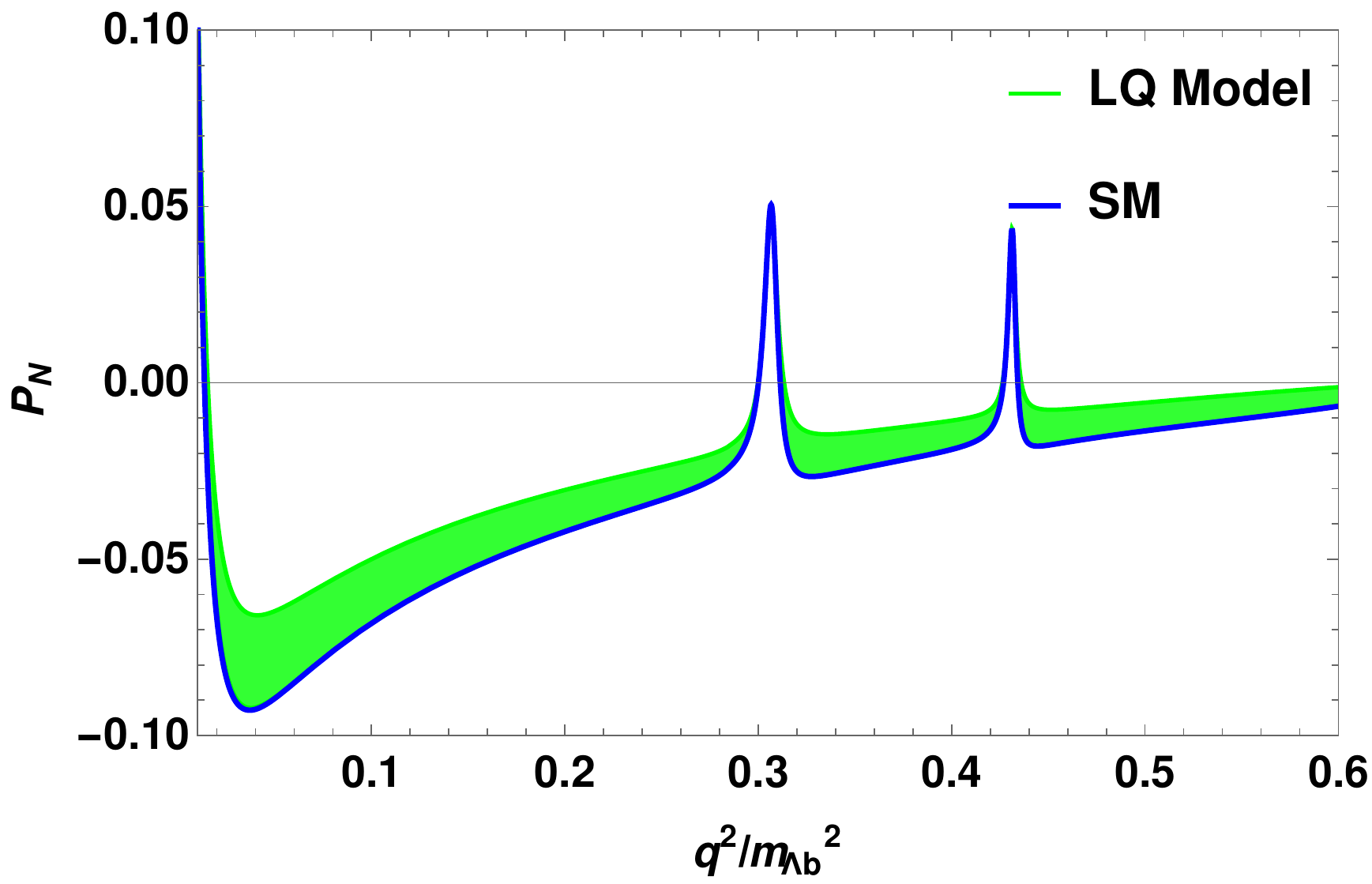}
\caption{The polarization plots of $\Lambda_b \rightarrow \Lambda \mu^+ \mu^-$ process for $X=(3, 2, 1/6)$ LQ exchange. }
\end{figure}
%%%%%%%%%%%%%%%%%%%%%%%%%%%%%%%%%%%%%%%%%%%%%%%%%%%%%%%%%%%%%%%%%%%%%%%%%%%%
\begin{figure}[h]
\centering
\includegraphics[scale=0.45]{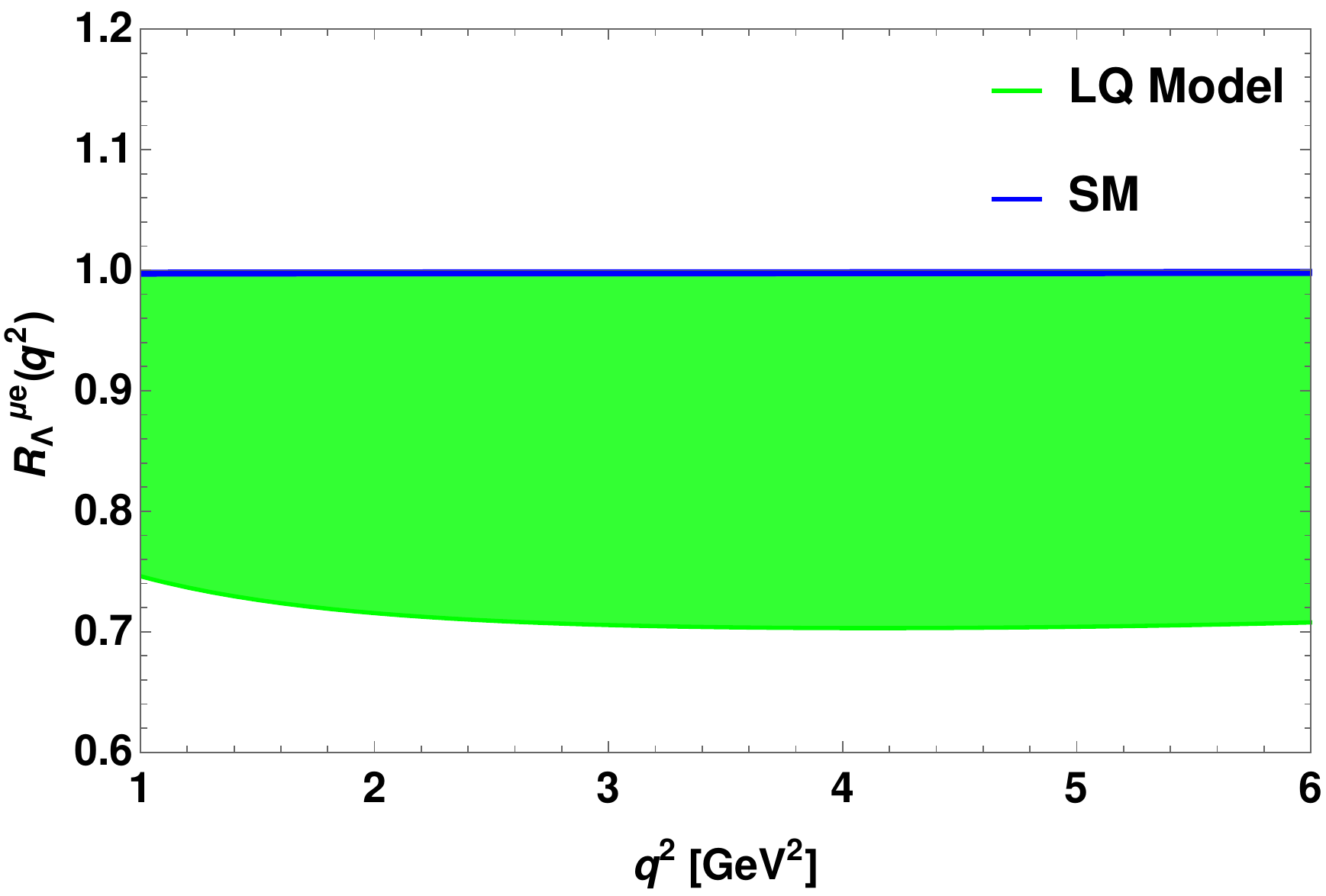}
\quad
\includegraphics[scale=0.45]{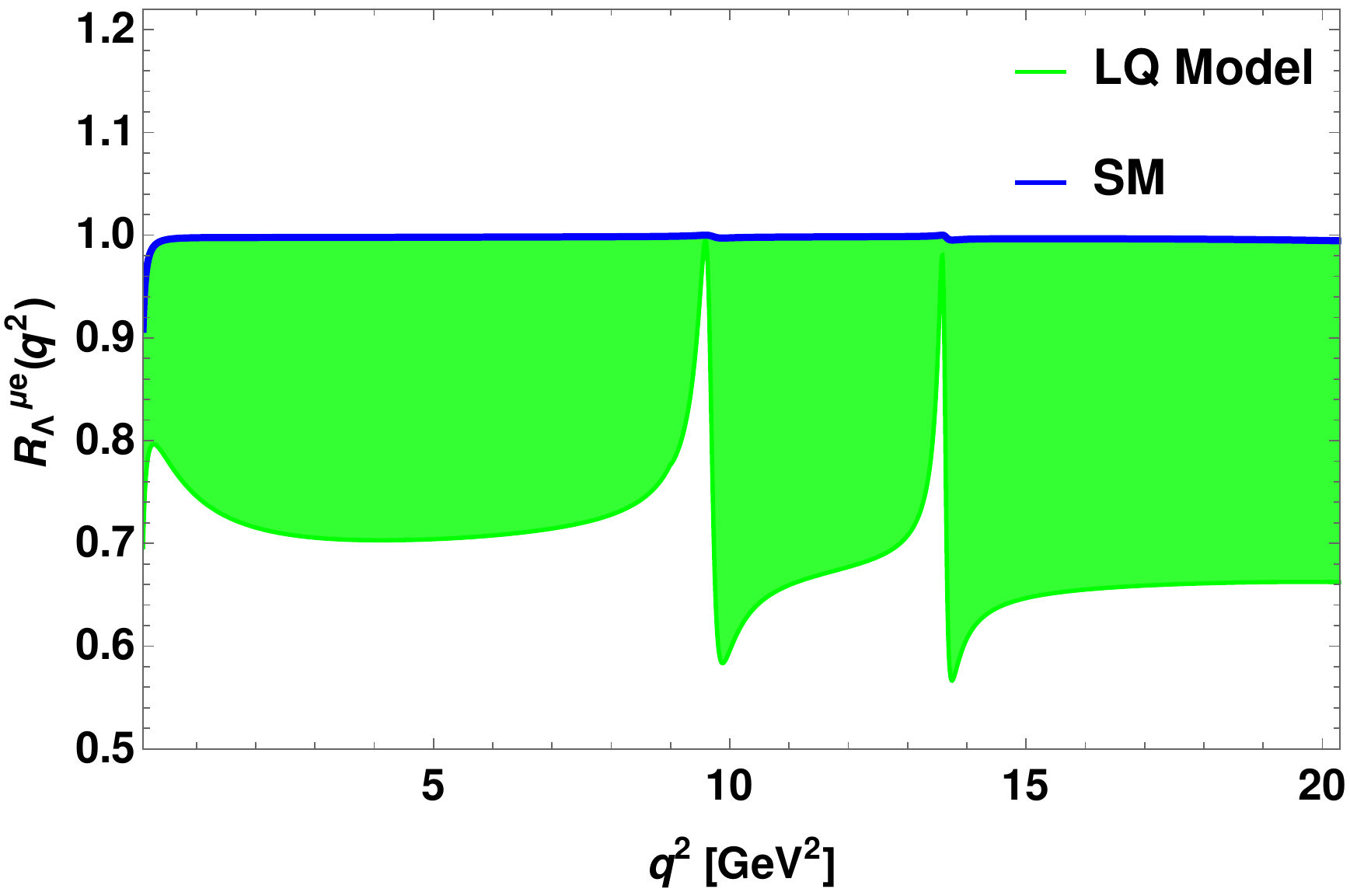}
\quad
\includegraphics[scale=0.45]{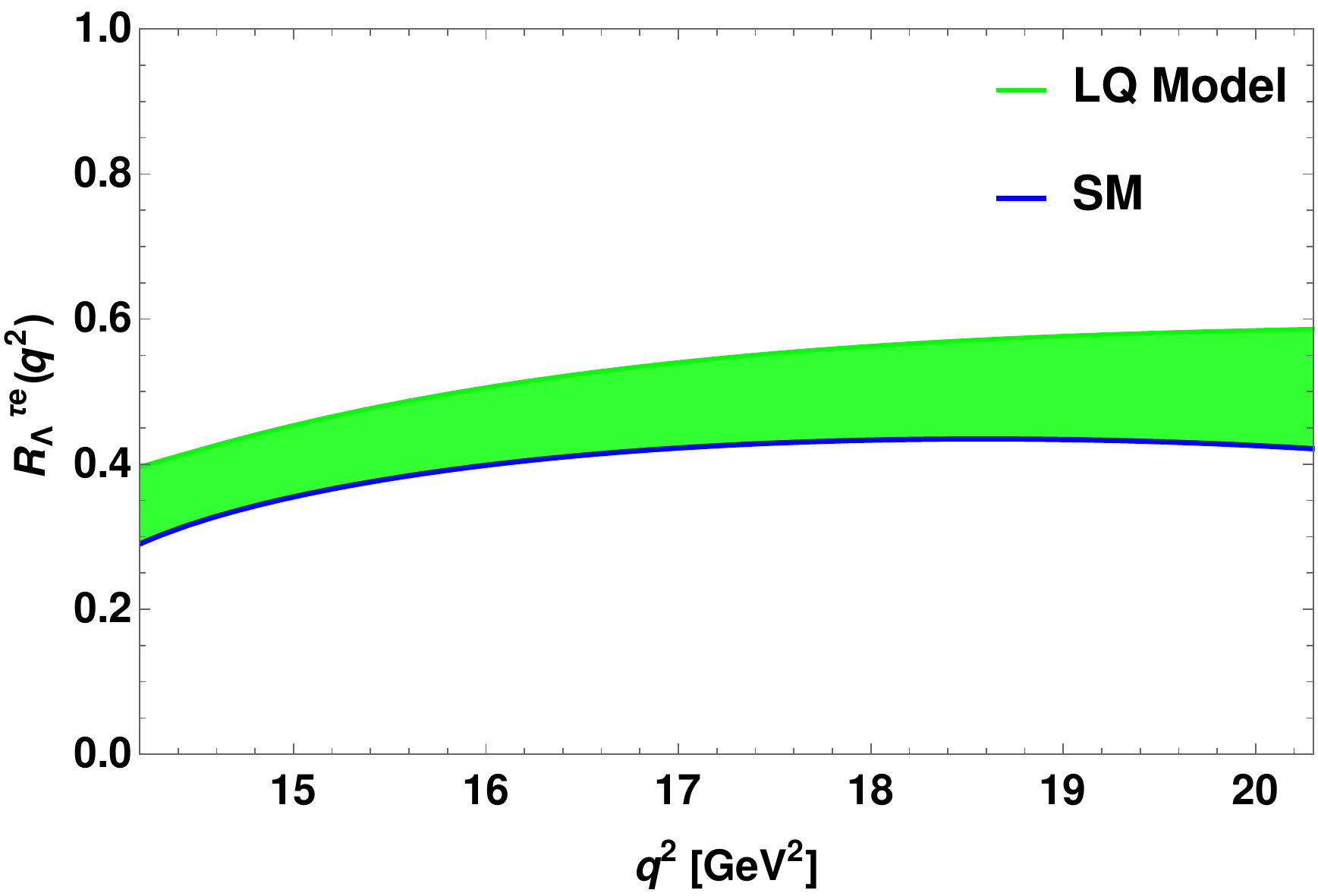}
\quad
\includegraphics[scale=0.45]{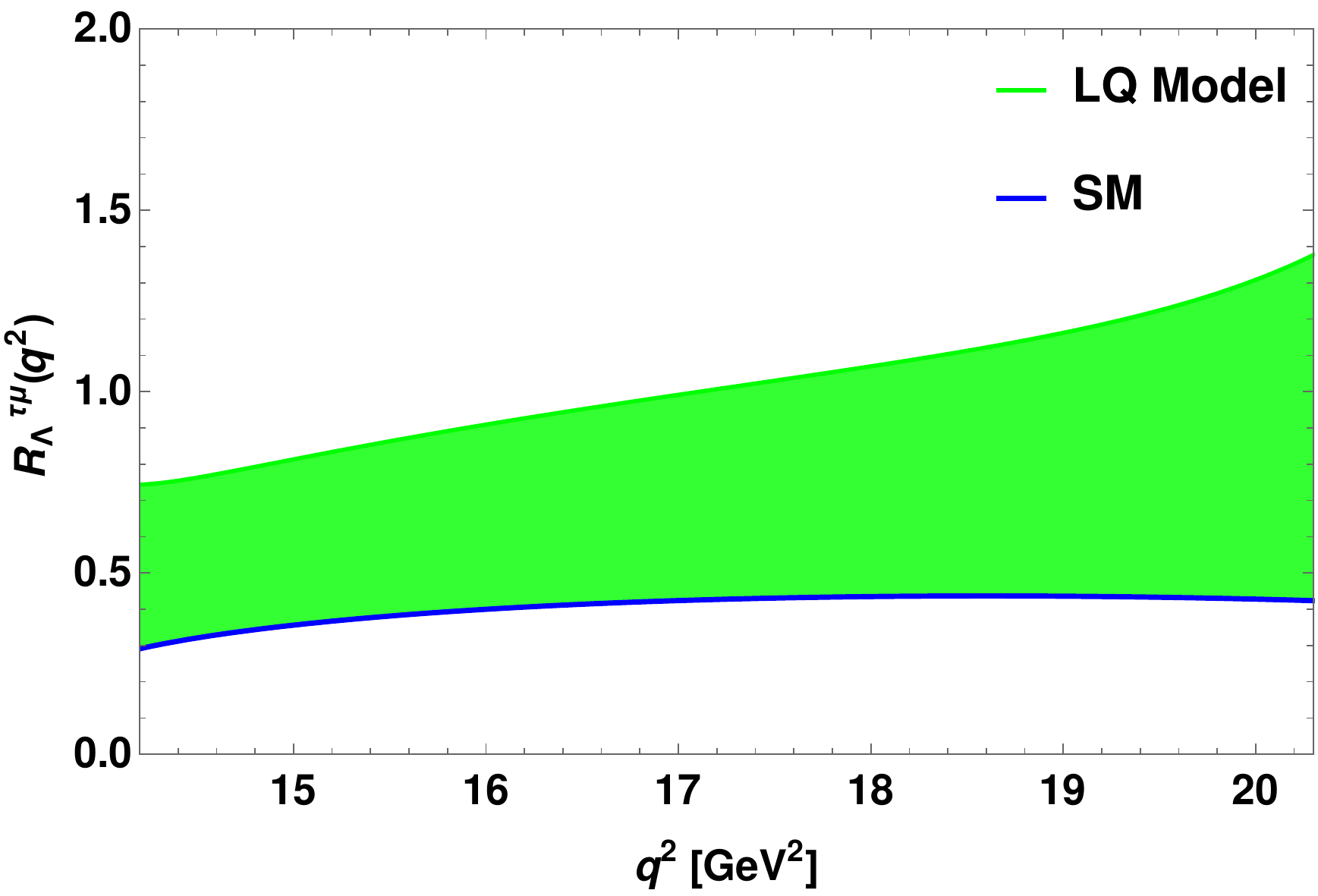}
\caption{The variation of  lepton universality violation $R_{\Lambda}^{\mu e}$ (top-right panel),  $R_{\Lambda}^{\tau e}$ (bottom-left panel) and $R_{\Lambda}^{\tau \mu}$ (bottom-right panel) with respect to $q^2$ for $X=(3, 2, 7/6)$ LQ exchange. Here $R_{\Lambda}^{\mu e}$ (top-left panel) shows the non-universality in the low $q^2 \in [1, 6]$ region.}
\end{figure}
%%%%%%%%%%%%%%%%%%%%%%%%%%%%%%%%%%%%%%%%%%%%%%%%%%%%%%%%%%%%%%%%
%%%%%%%%%%%
%%%%%%%%%%%%%%%%%%%%%%%%%%%%%%%%%%%%%%%%%%%%%%%%%%%%%%%%%%%%%%%%%%%%%%%%%%%%
\begin{figure}[h]
\centering
\includegraphics[scale=0.45]{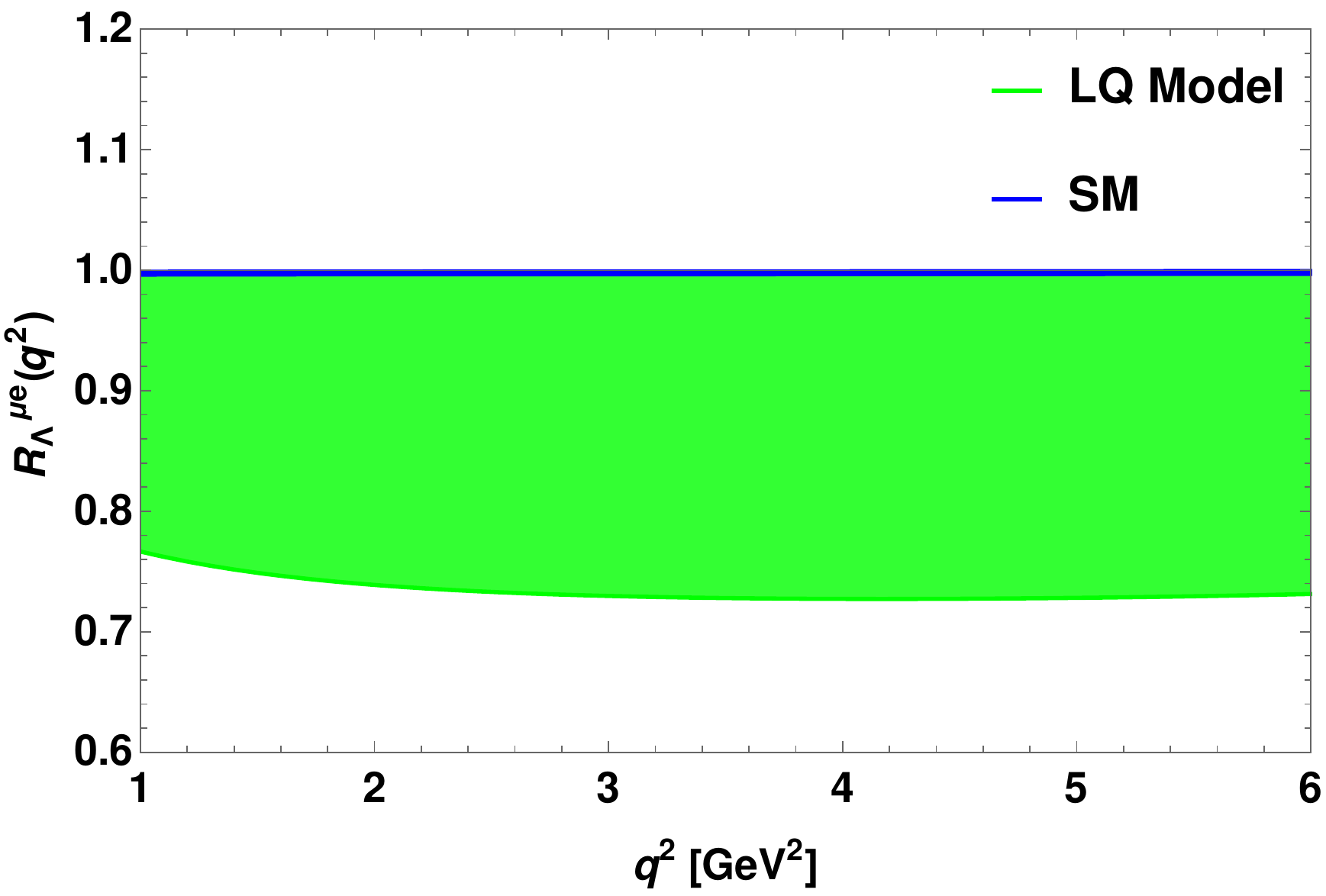}
\quad
\includegraphics[scale=0.45]{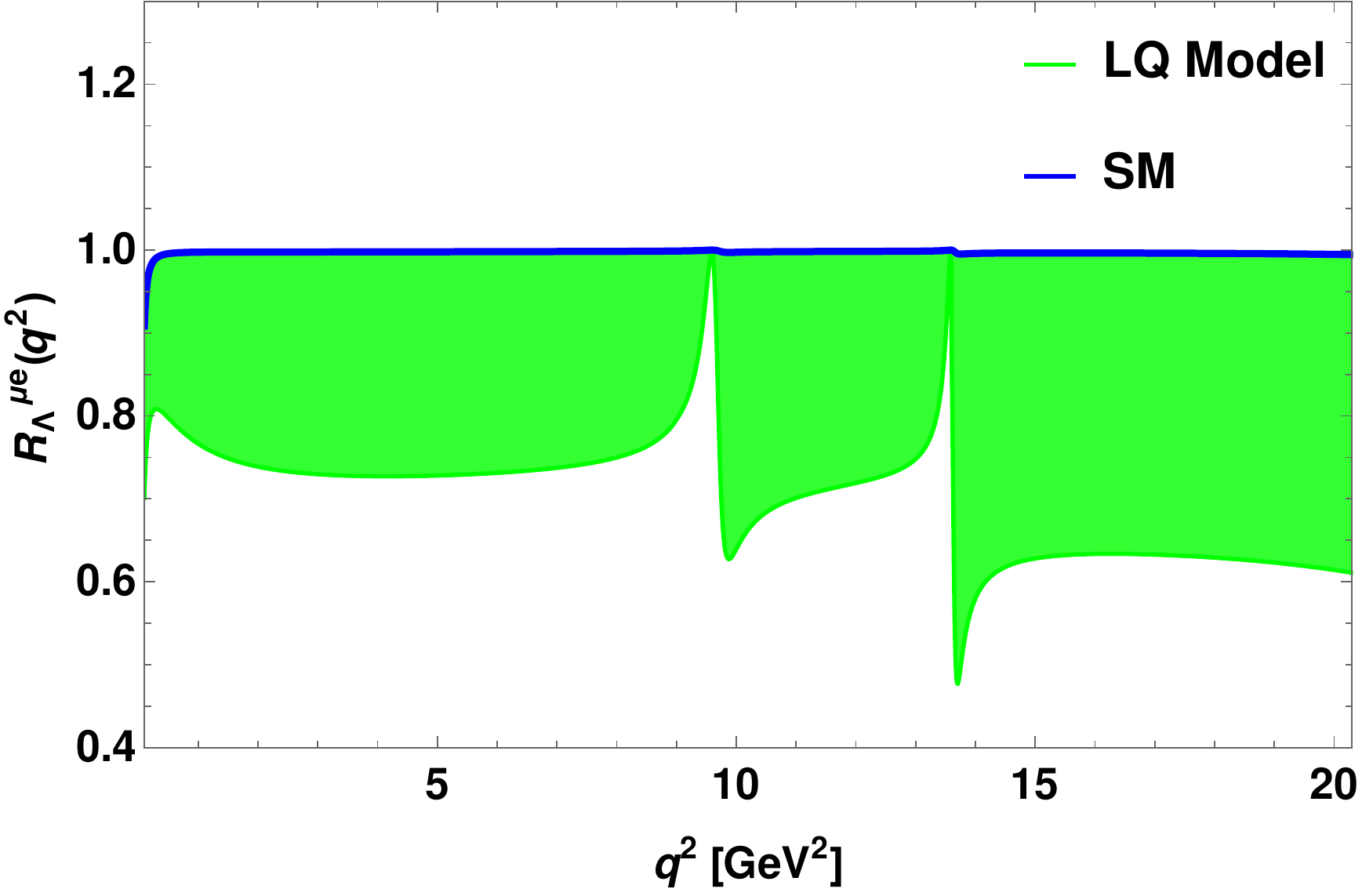}
\quad
\includegraphics[scale=0.45]{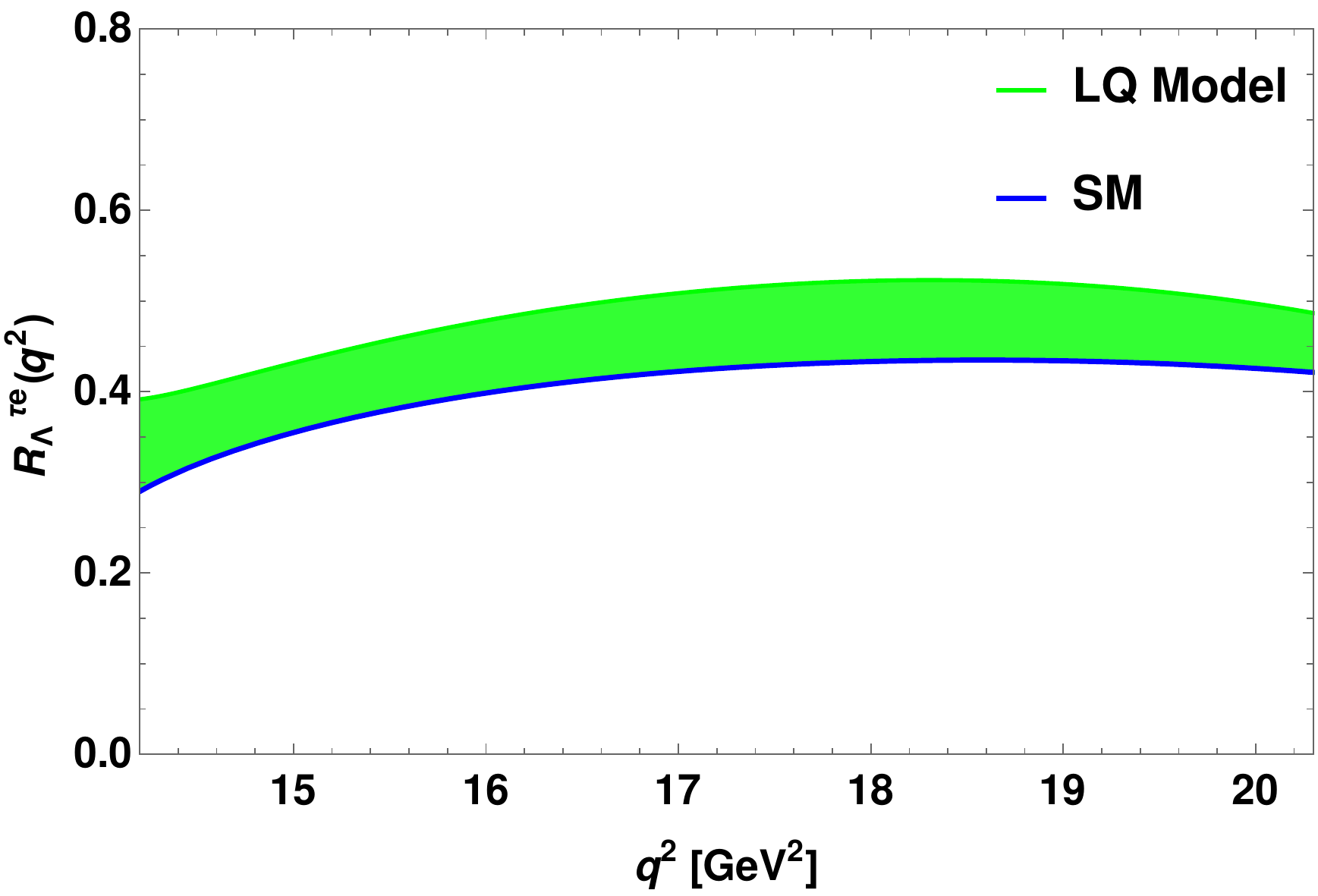}
\quad
\includegraphics[scale=0.45]{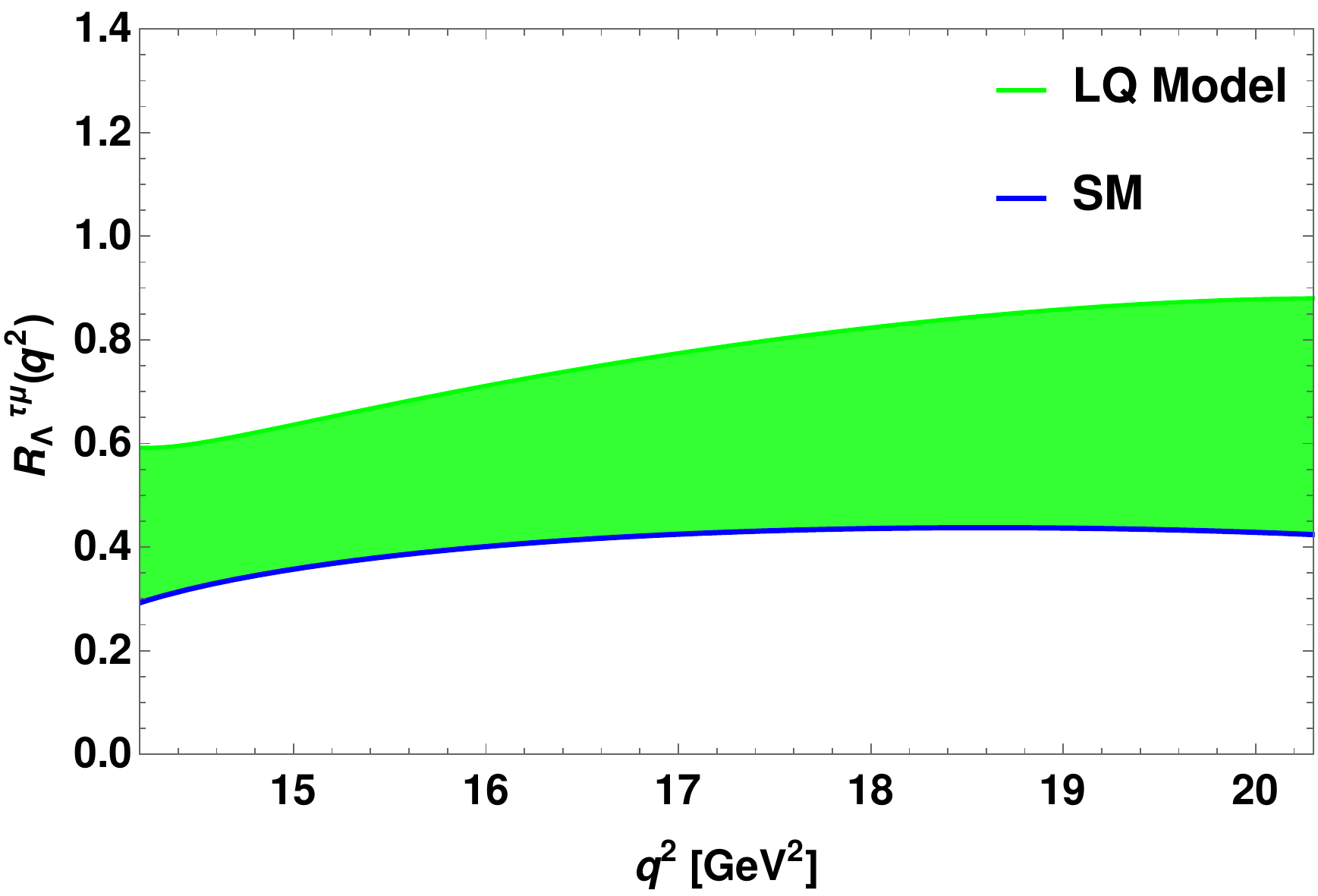}
\caption{Same as Fig.7 for $X=(3, 2, 1/6)$ LQ exchange.}
\end{figure}
%%%%%%%%%%%%%%%%%%%%%%%%%%%%%%%%%%%%%%%%%%%%%%%%%%%%%%%%%%%%%%
 %%%%%%%%%%%%%%%%%%%%%%%%%%%%%%%%%%%%%%%%%%%%%%%%%%%%%%%%%%%%%%%%%%%%%%%%%%%%%%%%%%%%%%%

\begin{table}[htb]
\begin{center}
\caption{The predicted  integrated values of the branching ratio, forward-backward asymmetry, lepton polarization asymmetry  and the lepton non-universality with respect to their respective $q^2$ range  for the $\Lambda_b \rightarrow \Lambda \mu(\tau)^+ \mu(\tau)^-$ processes in the SM and the LQ model. }
\begin{tabular}{|c | c | c| c|}
\hline
 Observables & SM prediction & Values in $Y=7/6$ LQ model & Values in $Y=1/6$ LQ model \\
 \hline
 \hline
 Br($\Lambda_b \rightarrow \Lambda  e^+ e^-$) & $(1.168 \pm 0.134) \times 10^{-6}$ & $(1.168 -1.91)\times 10^{-6}$ & $(1.168-2.13)\times 10^{-6}$\\
 
 Br($\Lambda_b \rightarrow \Lambda  \mu^+ \mu^-$) & $(1.165 \pm 0.132)\times 10^{-6}$ & $(1.165-1.37)\times 10^{-6}$ &  $(1.165-1.52)\times 10^{-6}$ \\
 
$\langle A_{FB}^\mu\rangle$ & $-0.567$ & $-0.567 \to -0.446$ & $-0.567 \to -0.54$\\

$\langle P_L^\mu\rangle$ & $0.34$ & $0.3-0.34$ & $0.24 -0.34$ \\

$\langle P_T^\mu \rangle$ & $-4.5\times 10^{-4}$ & $-(4.5 \to 2.87)  \times 10^{-4}$ & $-(0.45 \to 3.26)  \times 10^{-3}$\\

$\langle P_N^\mu \rangle$ & $-0.0192$ & $-0.0192 \to -0.013$ & $-0.0192 \to -0.012$ \\

\hline

Br($\Lambda_b \rightarrow \Lambda  \tau^+ \tau^-$) & $(2.13 \pm 0.215)\times 10^{-7}$ & $(2.13-4.38)\times 10^{-7}$ & $(2.13-8.32)\times 10^{-7}$ \\

$\langle A_{FB}^\tau\rangle$ & $-0.38$	& $-0.38 \to 3.2 \times 10^{-3}$ & $-0.38 \to 7.68 \times 10^{-2}$\\

$\langle P_L^\tau\rangle$ & $0.075$ & $0.047-0.075$ & $6.3 \times 10^{-3} - 0.075$ \\

$\langle P_T^\tau \rangle$ & $-2.3 \times 10^{-3}$ & $-(7 \to 2.3) \times 10^{-3}$& $ (- 0.23 \to 2.0)\times 10^{-2}$\\

$\langle P_N^\tau \rangle$ & $-0.05$ &$-0.05 \to 8.1 \times 10^{-3}$  & $-0.05\to 0.0316$ \\

\hline

$\langle R_{\Lambda_b}^{\mu e}\rangle$ & $0.997$ & $0.67- 0.997$ & $0.68 - 0.997$\\

$\langle R_{\Lambda_b}^{\mu e}\rangle_{[q^2 \in (1,6)]}$ & $0.998$	& $0.71- 0.998$ & $0.74 - 0.998$\\

 \hline
\end{tabular}
\end{center}
\end{table}
%%%%%%%%%%%%%%%%%%%%%%%%%%%%%%%%%%%%%%%%%%%%%%%%%%%%
\section{lepton flavour violating $\Lambda_b \rightarrow \Lambda l_i^- l_j^+$ decays}
%%%%%%%%%%%%%%%%%%%%%%%%%%%%%%%%%%%%%%%%%%%%%%%%%%%

In this section, we will compute the branching ratios of 
lepton flavour violating (LFV) $\Lambda_b$ decays mediating through the exchange of scalar leptoquarks.  The LFV decay processes are extremely rare in the SM as they are either two-loop suppressed with tiny neutrino masses in one of the loop or proceed through box diagram (which is also 
highly suppressed due to tiny neutrino mass). However,  they can occur at  tree level in the  LQ model and are expected to have significantly large branching fractions. 
The observation of neutrino oscillation has provided unambiguous evidence for lepton flavour violation in the neutral lepton sector which in turn provides motivation to explore other LFV transitions such as $l_i \to l_j \gamma$, 
$l_i^- \to l_j^- l_k^+ {l}_k^-$, $B \to l_i^\pm l_j^\mp$ etc.  Though there is no direct experimental evidence for such processes, but there exists experimental upper bounds  on some of these modes. The  LFV decays in the $B$ meson  and in the charged lepton sector have been widely investigated in the literature \cite{lfv, mohanta2, mohanta4}. Therefore, it is interesting to see whether LFV decays could be observed in $\Lambda_b$ decays also.

As discussed earlier, these processes occur at tree level due to the exchange of scalar
leptoquarks. 
In the leptoquark model the effective Hamiltonian for $b \to s l_i^- l_j^+$ LFV process  is given as \cite{mohanta2, mohanta4}
\bea
{\cal{H}}_{LQ} = G_{LQ} \left( \bar{s}\gamma^\mu P_L b \right) (\bar{l}_i\gamma_\mu (1+ \gamma_5) l_j) ,
\label{lfv-ham}
\eea 
where the  coefficient $G_{LQ}$ is
\be
G_{LQ} = \frac{\lambda^{i3} {\lambda^{j2}}^*}{8M_Y^2}\;.
\ee
Using the form factors given in the Appendix A, the amplitude for the LFV $\Lambda_b \to \Lambda l_i^- l_j^+$ decay is given by 
\bea
{\cal M}(\lb \to \ll l_i^- l_j^+) &=& G_{LQ} \Biggr[ \left( \bar{l}_i \gamma_\mu (1+\gamma_5) l_j\right) \Big\{
\bar u_\ll \Big(\gamma^\mu (A_1' P_R+B_1' P_L) \Big)u_{\lb} \nn\\
&+& 
\bar u_\ll i\sigma_{\mu \nu} q^\nu  (A_2' P_R+B_2' P_L) u_{\lb}
+ q^\mu \bar u_\ll (A_3' P_R +B_3' P_L) u_{\lb} \Big\}\Biggr]\;.\label{e1}
\eea
 The coefficients $A_k'$ and $B_k'$ in (\ref{e1}) are related to the
 form factors through 
\bea 
A_k' = \frac{f_k-g_k}{2} ~~~~ {\rm and} ~~~~~ B_k' = \frac{f_k+g_k}{2}, ~~~~k=1,2,3.
\eea
Now using this transition amplitude, the branching ratio for the
$\Lambda_b \to \Lambda l_i^- l_j^+$  process is given as
\bea
\frac{d^2\Gamma}{d \hat s d \cos\theta} = \frac{|G_{LQ}|^2}{2^6 \pi^3}  m_{\Lambda_b}^5 \frac{\sqrt{\lambda_1 \lambda_2}}{\hat s} I(\hat s),
\eea
where 
\bea
I(\hat s)= I_0(\hat s) +I_1(\hat s) \cos\theta +I_2(\hat s) \cos^2\theta,
\eea
with
\bea
I_0(\hat s) &= &\frac{1}{4}\left(|A_1'|^2+|B_1'|^2+m_{\lb}^2 \hat s (|A_2'|^2+|B_2'|^2) \right) \left[(1-r)^2-\hat s^2\right] \nn \\ &-& 2\sqrt{r} \hat s\left( 1-\frac{m_i^2+m_j^2}{q^2}\right) \left( Re (A_1' B_1'^*)+m_{\lb}^2 \hat s Re( A_2' B_2'^*) \right) \nn \\ &-&  Re\left( A_2' B_2'^*\right) \sqrt{r} \left[-\frac{\left(m_i^2-m_j^2 \right)^2}{m_{\Lambda b}^2}+\hat s \left( m_i^2+m_j^2 \right) \right]  \nn \\
 &+& 2m_{\lb} \hat s \left( 1-\frac{(m_i^2+m_j^2)}{2q^2}\right) 
 \Big[ \left(Re( A_1' A_2'^*) +Re( B_1' B_2'^*) \right) \sqrt{r} (1-t)\nn\\
&-& \Big(Re( A_1' B_2'^*) +Re (B_1' A_2'^*) \Big) (t-r) \Big] \nn \\ &+& \frac{(m_i^2+m_j^2)}{m_{\lb}} \Big[\Big(Re (A_1' A_3'^*) +Re (B_1' B_3'^*) \Big) \sqrt{r} (1-t) +\Big(Re (A_1' B_3'^*) +Re (B_1' A_3'^*) \Big) (t-r) \Big] \nn \\ &+& \left[\hat s(m_i^2+m_j^2) -\frac{(m_i^2-m_j^2)^2}{m_{\lb}^2} \right] \left[\frac{t}{2}(|A_3'|^2+|B_3'|^2) -\sqrt{r} Re( A_3' B_3'^* )\right], 
\eea
\bea
I_1(\hat s) &=&\frac{\sqrt{\lambda_1 \lambda_2}}{\hat s} \Big[ -\frac{1}{2} \hat s (|A_1'|^2-|B_1'|^2) + (m_j^2-m_i^2) (1-t-\frac{\hat s}{2}) (|A_2'|^2+|B_2'|^2) \nn \hspace{3.5cm} \\ &+& \frac{1}{2} \frac{m_j^2-m_i^2}{m_{\lb}} \left[ \sqrt{r}\left(Re (A_1' A_2'^*) +Re( B_1' B_2'^*) \right) - \left(Re  (A_1' B_2'^*) +Re (B_1' A_2'^*) \right) \right] \nn \hspace{3.5cm} \\ &+& \frac{1}{2} \frac{m_j^2-m_i^2}{m_{\lb}} \left[ \sqrt{r}\left(Re (A_1' A_3'^*) +Re (B_1' B_3'^*) \right) + \left(Re ( A_1' B_3'^*) +Re (B_1' A_3'^*) \right) \right] \nn \hspace{3.5cm} \\   &+& \frac{\hat s}{2} (m_i^2-m_j^2) \left(Re (A_2' A_3'^*) +Re (B_2' B_3'^*) \right)  \Big],
\eea
and
\bea
I_2(\hat s) = \frac{\lambda_1 \lambda_2}{\hat s^2} \left[-\frac{1}{4}\left(|A_1'|^2+|B_1'|^2-m_{\lb}^2 \hat s (|A_2'|^2 +|B_2'| ^2) \right) \right].\hspace{5cm}
\eea
Here, $\lambda_{1}=\lambda$ (as defined in section III), $\lambda_2=\hat{m}_i^4+\hat{m}_j^4+ \hat s^2-2 \left(\hat{m}_i^2 \hat{m}_j^2 +\hat{m}_i^2 \hat s +\hat{m}_j^2 \hat s \right)$, and $t=(1+r-\hat s)/2$. The full kinematically accessible physical range for these processes is given by
\bea
(m_i+m_j)^2 \leq q^2 \leq (m_{\Lambda_b}-m_\Lambda)^2.
\eea
As there is no intermediate particle in the SM which can decay into two leptons of different  flavours, so in comparison with the  $\Lambda_b \to \Lambda l^+ l^-$ processes, LFV decays have no long distance QCD contributions and dominant charmonium resonance background.
The required  input values for numerical evaluation  are taken from \cite{pdg} and the values of the $q^2$ dependent form factors are taken  from LCSR approach \cite{LCSR}.  To determine the values of various LQ couplings, which are involved in the LFV decays, we use the following assumptions. As we know that the expansion parameter of the CKM matrix in the Wolfenstein parametrization ($\lambda$), can be related to the down type quark masses as $\lambda \sim (m_d/m_s)^{1/2}$ in the quark sector, while in the lepton sector one can have the same order for $\lambda$ with the relation $\lambda \sim (m_{l_i}/m_{l_j})^{1/4}$. Hence,  for other required leptoquark coupling, we  assume that the coupling between different generation of quarks and leptons follow the simple scaling laws, i.e., $\lambda^{ij} = (m_i/m_j)^{1/4} \lambda^{ii}$ with $j>i$. 
Thus, using the values of  the leptoquark coupling as  given in Table I, one can  obtain the bound on required LQ couplings involved in LFV decays.  Using these values we  plot the variation of branching ratio of LFV decays such as $\Lambda_b \rightarrow \Lambda \mu^- e^+$ (top left panel), $\Lambda_b \rightarrow \Lambda \tau^- e^+$ (top right panel) and $\Lambda_b \rightarrow \Lambda \tau^- \mu^+$ (lower panel) with respect to $q^2$ in Fig. 9 and the predicted upper limits of the branching ratios are  given in Table IV. So far there is no experimental evidence on the  LFV $\Lambda_b$ decays. However, since the predicted branching ratios are
${\cal O}(10^{-9})$, they can be searched at  LHCb and  exploration/observation of these modes would definitely shed some light in the leptoquark  scenarios.
%%%%%%%%%%%%%%%%%%%%%%%%%%%%%%%%%%%%%%%%%%%%%%%%%%%%%%%%%%%%%%%%%%%%%%%%%%%%%%
\begin{figure}[h]
\centering
\includegraphics[scale=0.45]{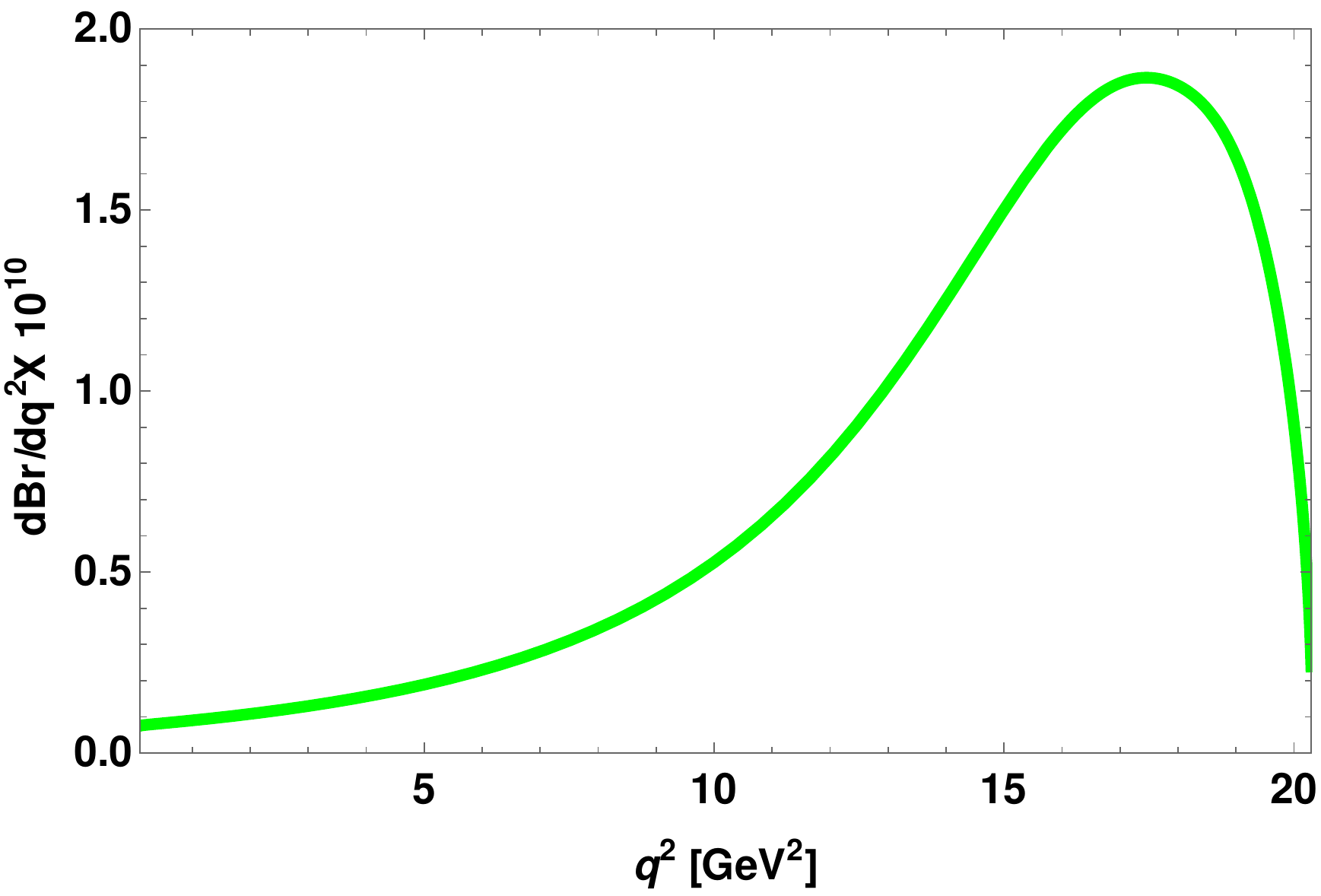}
\quad
\includegraphics[scale=0.45]{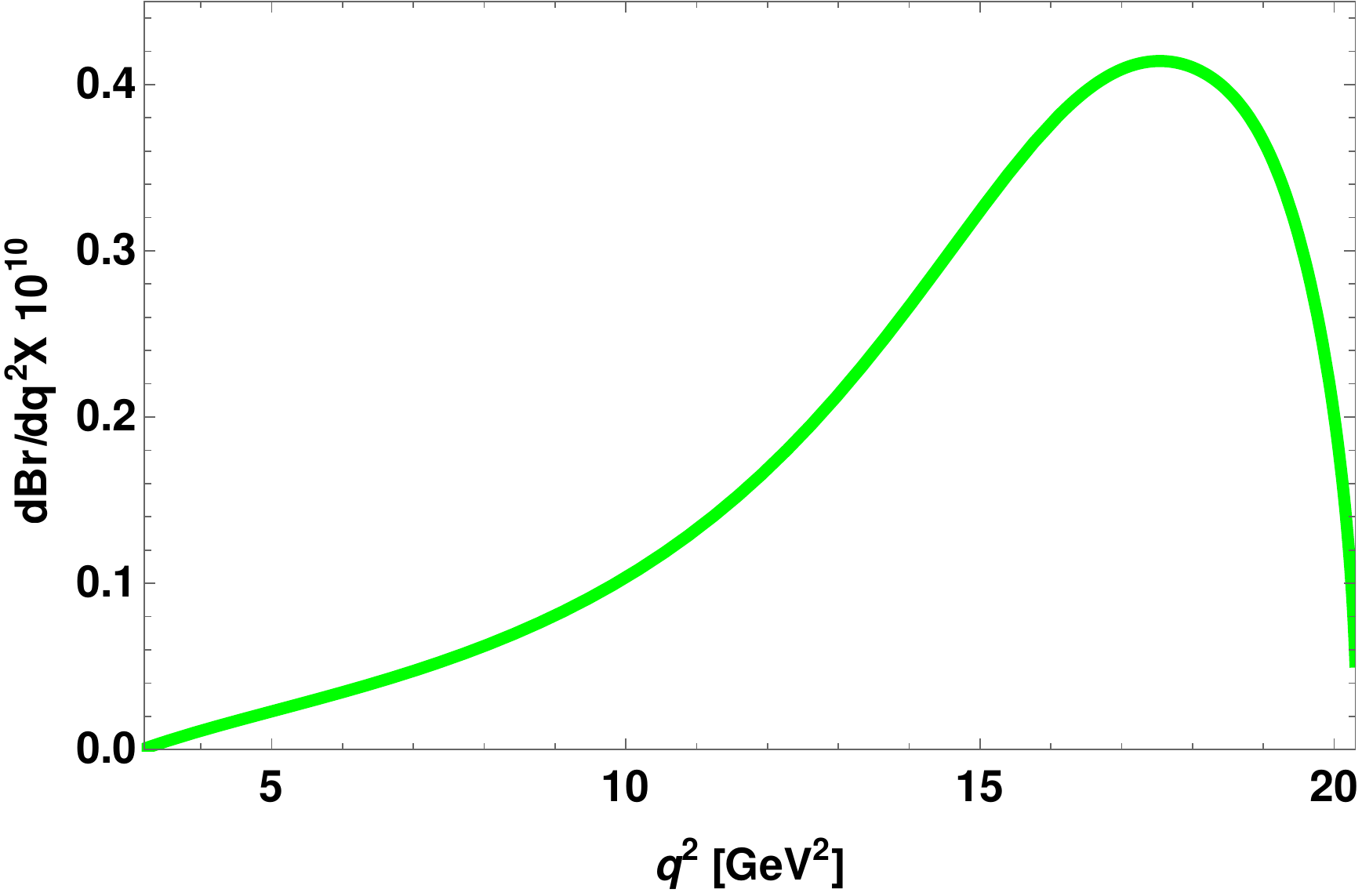}
\quad
\includegraphics[scale=0.45]{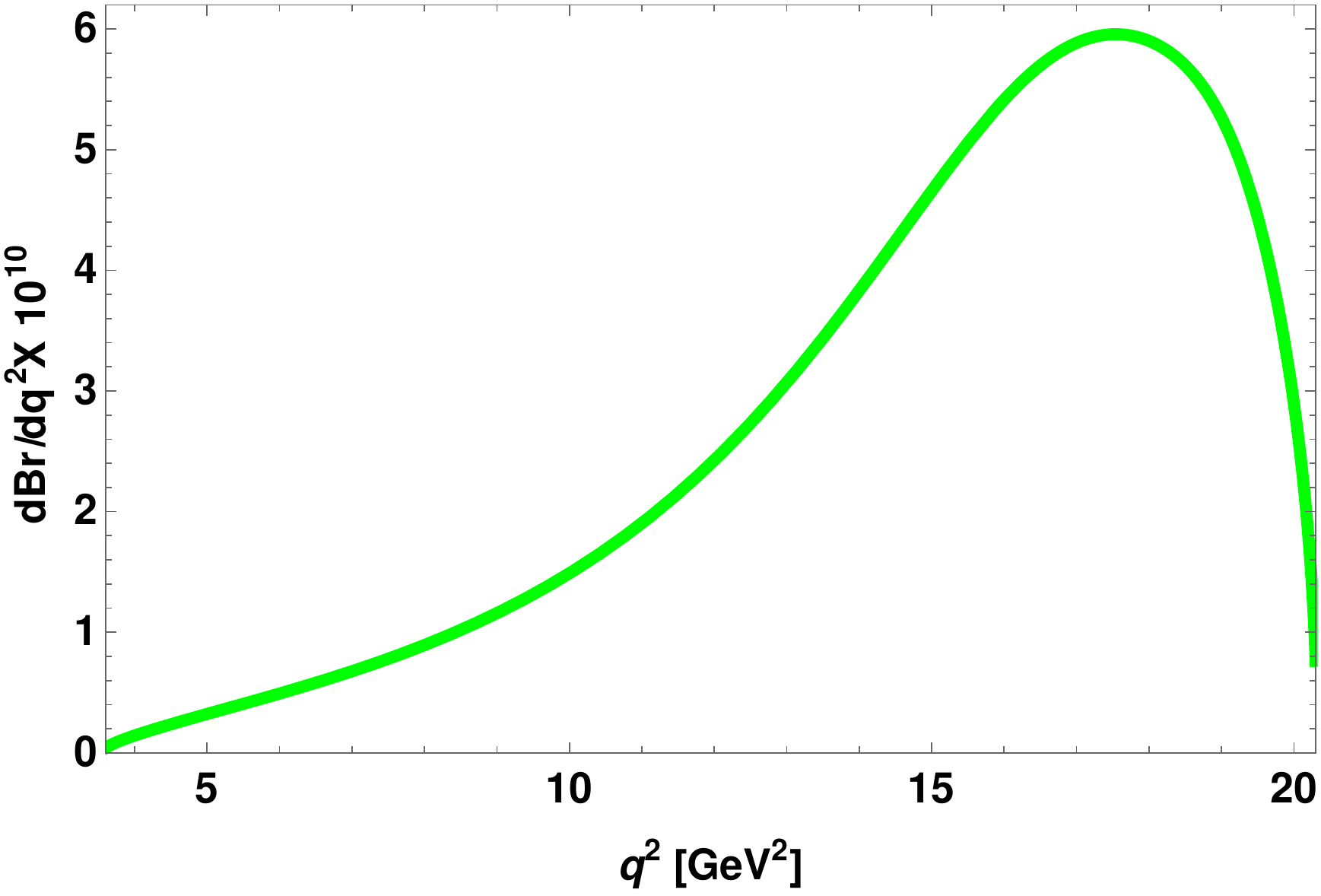}
\caption{The variation of  branching ratio of LFV  $\Lambda_b \rightarrow \Lambda \mu^- e^+$ (left panel), $\Lambda_b \rightarrow \Lambda \tau^- e^+$ (right panel)  and $\Lambda_b \rightarrow \Lambda \tau^- \mu^+$ (bottom panel) processes  with respect to $q^2$ in the $X=(3, 2, 7/6)$ leptoquark model. Here the required leptoquark couplings are computed by using the scaling ansatz $\lambda^{ij} = (m_i/m_j)^{1/4} \lambda^{ii}$.  }
\end{figure}
%%%%%%%%%%%%%%%%%%%%%%%%%%%%%%%%%%%%%%%%%%%%%%%%%%%%%%%%%%%%%%%%%%%%%%%%%%%%%%
\begin{table}[h]
\caption{The predicted upper limits of the  branching ratios, which are obtained using the upper limits of the LQ couplings,   of LFV $\Lambda_b \rightarrow \Lambda l_i^- l_j^+$  processes, $l = e, \mu, \tau$ in the $X=(3, 2, 7/6)$ leptoquark model. Also the required leptoquark couplings are computed by using the scaling ansatz $\lambda^{ij} = (m_i/m_j)^{1/4} \lambda^{ii}$.}
\begin{center}
\begin{tabular}{| c | c |}
\hline
 Decay process  & Predicted branching ratio   \\
 
 \hline
 \hline
$\Lambda_b \rightarrow \Lambda  \mu^- e^+$   &  $ 1.56 \times 10^{-9}$\\
$\Lambda_b \rightarrow \Lambda  \tau^- e^+$   &  $ 3.2 \times 10^{-10}$\\
$\Lambda_b \rightarrow \Lambda  \tau^- \mu^+$  &  $ 4.6 \times 10^{-9}$\\
 \hline
\end{tabular}
\end{center}
\end{table}
%%%%%%%%%%%%%%%%%%%%%%%%%%%%%%%%%%%%%%%%%%%%%%%%%%%%%%%%%%%%%%%%%%%%%%%%%%%%%%
\section{conclusion}
In this paper, we have studied the rare semileptonic $\Lambda_b \to \Lambda l^+ l^-$, $l = e,\mu, \tau$ baryonic decays in the scalar leptoquark model. The leptoquark parameter space has been constrained  using the experimental limits on the branching ratios of the two body leptonic decays $B_s \to l^+ l^-$. We have computed the  branching ratios, the forward-backward  and lepton polarization asymmetries $(P_{L, T, N})$  using the new leptoquark couplings. We have shown explicitly  the results for both the relevant $X=(3, 2, 7/6)$ and $X=(3, 2, 1/6)$ leptoquark models. The zero-position of the forward-backward asymmetry is found to be insensitive  to the additional leptoquark effect. These models also give negligible contribution to the transverse polarization asymmetry.   In addition, we also estimated the  lepton universality violation parameters in these decays analogous to $R_K$ in  $B \to K l^+ l^-$ process. The lepton flavour violating $\Lambda_b$ decays are also studied and the predicted  upper limits on these branching ratios  are
found to be ${\cal O}(10^{-10}-10^{-9})$, which  could be searched in the LHCb  experiment.

%%%%%%%%%%%%%%%%%%%%%%%%%%%%%%%%%%%%%%%%%%%%%%%%%%%%%%%%%%%%%%%%%%%%%%%%%%%%%%%%%%%%%%%%%
%%
\appendix
\section{ Form Factors in $\Lambda_b \to \Lambda$ transition}
The transition form factors for  $\Lambda_b (P) \to \Lambda (p')$  decays can be parameterized as \cite{chen1, mannel}
\bea
&&\langle \Lambda (p') |\bar{s}\gamma_\mu b |\Lambda_b (P) \rangle = f_1 \bar{u}_\Lambda \gamma_\mu {u}_{\Lambda_b} +  f_2 \bar{u}_\Lambda i\sigma_{\mu \nu}q^\nu {u}_{\Lambda_b} +  f_3 q_\mu \bar{u}_\Lambda {u}_{\Lambda_b}, \nn \\
&&\langle \Lambda (p')|\bar{s}\gamma_\mu \gamma_5 b |\Lambda_b(P) \rangle = g_1 \bar{u}_\Lambda \gamma_\mu \gamma_5 {u}_{\Lambda_b} +  g_2 \bar{u}_\Lambda i\sigma_{\mu \nu}q^\nu \gamma_5 {u}_{\Lambda_b} +  g_3 q_\mu \bar{u}_\Lambda \gamma_5 {u}_{\Lambda_b}, \nn  \\
&&\langle \Lambda(p')| \bar{s} i\sigma_{\mu \nu} b |\Lambda_b (P) \rangle = f_T \bar{u}_\Lambda i\sigma_{\mu \nu} {u}_{\Lambda_b} +  f_T^V \bar{u}_\Lambda \left(\gamma_\mu q_\nu - \gamma_\nu q_\mu \right) {u}_{\Lambda_b} +  f_T^S \left(P_\mu q_\nu - P_\nu q_\mu \right) \bar{u}_\Lambda {u}_{\Lambda_b}, \nn  \\
&&\langle \Lambda(p')| \bar{s} i\sigma_{\mu \nu} \gamma_5 b |\Lambda_b(P) \rangle = g_T \bar{u}_\Lambda i\sigma_{\mu \nu} \gamma_5 {u}_{\Lambda_b} +  g_T^V \bar{u}_\Lambda \left(\gamma_\mu q_\nu - \gamma_\nu q_\mu \right) \gamma_5 {u}_{\Lambda_b}\nn\\
&&\hspace*{1.6truein} +  g_T^S \left(P_\mu q_\nu - P_\nu q_\mu \right) \bar{u}_\Lambda  \gamma_5{u}_{\Lambda_b}, 
\eea
 and  for dipole operators 
\begin{eqnarray}
\left\langle \Lambda(p') \right| \bar{s}i\sigma _{\mu \nu }q^{\nu }b\left|
\Lambda _{b}(P)\right\rangle &=&f_{1}^{T}\bar{u}_{\Lambda }\gamma _{\mu
}u_{\Lambda _{b}}+f_{2}^{T}\bar{u}_{\Lambda }i\sigma _{\mu \nu }q^{\nu
}u_{\Lambda _{b}}+f_{3}^{T}q_{\mu }\bar{u}_{\Lambda }u_{\Lambda _{b}},
\label{tcq} \\
\left\langle \Lambda (p') \right| \bar{s}i\sigma _{\mu \nu }q^{\nu }\gamma
_{5}b\left| \Lambda _{b} (P)\right\rangle &=&g_{1}^{T}\bar{u}_{\Lambda }\gamma
_{\mu }\gamma _{5}u_{\Lambda _{b}}+g_{2}^{T}\bar{u}_{\Lambda }i\sigma _{\mu
\nu }q^{\nu }\gamma _{5}u_{\Lambda _{b}}+g_{3}^{T}q_{\mu }\bar{u}_{\Lambda
}\gamma _{5}u_{\Lambda _{b}}.  \label{atcq}
\end{eqnarray}
with $q=P-p'$ and
\bea
f_{2}^{T} &=&f_{T}-f_{T}^{S}q^{2}\,,
\nn \\
f_{1}^{T} &=&\left[ f_{T}^{V}+f_{T}^{S}\left( M_{\Lambda }+M_{\Lambda
_{b}}\right) \right] q^{2}\,, \nn \\
f_{1}^{T} &=&-\frac{q^{2}}{\left( M_{\Lambda _{b}}-M_{\Lambda }\right) }%
f_{3}^{T}\, ,\nn \\
g_{2}^{T} &=&g_{T}-g_{T}^{S}q^{2}\,,\nn \\
g_{1}^{T} &=&\left[ g_{T}^{V}+g_{T}^{S}\left( M_{\Lambda }-M_{\Lambda
_{b}}\right) \right] q^{2}\,,\nn \\
g_{1}^{T} &=&\frac{q^{2}}{\left( M_{\Lambda _{b}}+M_{\Lambda }\right) }%
g_{3}^{T}\,.
\eea

\section{ Expressions for  ${\cal K}_{0, 1, 2}(\hat s)$
functions }
The complete expressions for ${\cal K}_{0, 1, 2}(\hat s)$ functions required to calculate the double differential decay rate is given by \cite{mohanta3}

\bea
{\cal K}_0(\hat s) &=& 32 m_l^2 m_{\lb}^2\s1(1+r -\s1)(|D_3|^2+|E_3|^2)\nn\\
&+&
64 m_l^2 m_{\lb}^3(1-r -\s1)Re(D_1^*E_3+D_3 E_1^*)
+64  m_{\lb}^2 \sqrt{r} (6 m_l^2-\s1 m_{\lb}^2)Re(D_1^*E_1)
\nn\\
&+&64 m_l^2 m_{\lb}^3 \sqrt{r} \Big(
2 m_{\lb} \s1 Re(D_3^*E_3)+(1-r+\s1)Re(D_1^* D_3+ E_1^* E_3)\Big)\nn\\
&+&32 m_{\lb}^2 (2 m_l^2+m_{\lb}^2 \s1)\Big((1-r +\s1)m_{\lb}
\sqrt{r}Re(A_1^*A_2+B_1^* B_2)\nn\\
&-& m_{\lb}(1-r-\s1) Re(A_1^* B_2+A_2^* B_1)-2 \sqrt{r}\Big[Re(A_1^* B_1)
+m_{\lb}^2 \s1 Re(A_2^* B_2)\Big]\Big)\nn\\
&+& 8 m_{\lb}^2\Big(4 m_l^2(1+r-\s1)+m_{\lb}^2[(1-r)^2-\s1^2]\Big)
\Big(|A_1|^2+|B_1|^2\Big)\nn\\
&+& 8 m_{\lb}^4\Big(4 m_l^2[\lambda+(1+r-\s1)\s1]+m_{\lb}^2
\s1[(1-r)^2-\s1^2]\Big)
\Big(|A_2|^2+|B_2|^2\Big)\nn\\
&-& 8 m_{\lb}^2\Big(4 m_l^2(1+r-\s1)-m_{\lb}^2[(1-r)^2-\s1^2]\Big)
\Big(|D_1|^2+|E_1|^2\Big)\nn\\
&+& 8 m_{\lb}^5 \s1 v_l^2 \Big(-8 m_{\lb} \s1 \sqrt{r}
Re(D_2^* E_2)+4 (1-r+\s1)\sqrt{r}Re(D_1^* D_2+E_1^* E_2)\nn\\
&-&4(1-r -\s1) Re(D_1^* E_2+D_2^* E_1)
+m_{\lb}[(1-r)^2-\s1^2]
\Big[|D_2|^2+|E_2|^2\Big]\Big)\;,
\eea
\bea
{\cal K}_1(\hat s) &=& -16  m_{\lb}^4\s1 v_l \sqrt{\lambda}
\Big\{ 2 Re(A_1^* D_1)-2Re(B_1^* E_1)\nn\\
&+& 2m_{\lb}
Re(B_1^* D_2-B_2^* D_1+A_2^* E_1-A_1^*E_2)\Big\}\nn\\
&+&32 m_{\lb}^5 \s1~ v_l \sqrt{\lambda} \Big\{
m_{\lb} (1-r)Re(A_2^* D_2 -B_2^* E_2)\nn\\
&+&
\sqrt{r} Re(A_2^* D_1+A_1^* D_2-B_2^*E_1-B_1^* E_2)\Big\}\;, \hspace{6cm}
\eea
and
\bea
{\cal K}_2(\hat s)&= & 8m_{\lb}^6 v_l^2~ \lambda \s1~ \Big ( 
(|A_2|^2+|B_2|^2+|D_2|^2+|E_2|^2\Big)\nn\\
&-&8m_{\lb}^4 v_l^2 ~\lambda~\Big(|A_1|^2+|B_1|^2+|D_1|^2+|E_1|^2\Big)\;. \hspace{5.5cm}
\eea
%%%%%%%%%%%%%%%%%%%%%%%%%%%%%%%%%%%%%%%%%%%%%%%%%%%%%%%%%%%%%%%%%%%%%%%%%%%%%%%

%%%%%%%%%%%%%%%%%%%%%%%%%%%%%%%%%%%%%%%%%%%%%%%%%%%%%%%%%%%%%%%%%%%%%%%%%%%%%%%%%%%
{\bf Acknowledgments} 

We would like to thank Science and Engineering Research Board (SERB) for financial support through grant No. SB/S2/HEP-017/2013.

%%%%%%%%%%%%%%%%%%%%%%%%%%%%%%%%%%%%%%%%%%%%%%%%%%%%%%%%%%%%%%%%%%%%%%%%%%%%%%%%%%%%%
%%%%%%%%%%%%%%%%%%%%%%%%%%%%%%%%%%%%%%%%%%%%%%%%%%%%%%%%%%%%%%%%%%%%%%%%%%%%%%%%%%%%


\begin{thebibliography}{60}

\bibitem{lhcb1}
  R. Aaij et al., [LHCb Collaboration],  
  Phys.\ Rev.\ Lett.\  {\bf 111}, 191801 (2013), 
  [arXiv:1308.1707 [hep-ex]].
 \bibitem{lhcb1a}
  R. Aaij et al., [LHCb Collaboration], JHEP \textbf{02} (2016) 104,  [arXiv:1512.04442].
 
 
 \bibitem{p5p}
  A.~Abdesselam {\it et al.} [Belle Collaboration],
  arXiv:1604.04042 [hep-ex].
  
\bibitem{lhcb2}
R. Aaij et al., [LHCb Collaboration], Phys. Rev. Lett. \textbf{113}, 151601 (2014) [arXiv:1406.6482].

\bibitem{lhcb3}
R. Aaij et al., [LHCb Collaboration], JHEP \textbf{1406}, 133 (2014) [arXiv:1403.8044].


\bibitem{lhcb4}
R. Aaij et al., [LHCb Collaboration], Phys. Rev. Lett. \textbf{111}, 191801 (2013) [arXiv:1308.1707].



  
\bibitem{lhcb5}
R. Aaij et al., [LHCb Collaboration], JHEP \textbf{1307}, 084 (2013) [arXiv:1305.2168].




\bibitem{pdg}
 K.A. Olive et al. (Particle Data Group), Chin. Phys. C, 38, 090001 (2014).



\bibitem{matias1}
S. Descotes-Genon, J. Matias, M. Ramon, J. Virto, JHEP \textbf{1301}, 048 (2013) [arXiv:1207.2753].

\bibitem{jager}
 S. Jager, J. Martin Camalich, JHEP \textbf{05}, 043 (2013) [arXiv:1212.2263]; S. Descotes-Genon, L. Hofer, J. Matias and J. Virto, JHEP \textbf{1412}, 125 (2014) [arXiv:1407.8526].
 
 \bibitem{jager15} S. Descotes-Genon, L. Hofer, J. Matias, J. Virto,
  JHEP {\bf 06}, 092 (2016) [arXiv:1510.04239].

 \bibitem{huber}
  T. Huber, T. Hurth and E. Lunghi, Nucl. Phys. B \textbf{802}, 40 (2008),[arXiv:0712.3009].
   
\bibitem{beaujean}
  F. Beaujean, C. Bobeth, and D. van Dyk, Eur. Phys. J. C \textbf{74}, 2897, (2014) [arXiv:1310.2478];
T. Hurth and F. Mahmoudi, JHEP \textbf{04}, 097 (2014) [arXiv:1312.5267]; W. Altmannshofer, S. Gori, M. Pospelov and I. Yavin, Phys. Rev. 
D \textbf{89}, 095033 (2014) [arXiv: 1403.1269]; G. Hiller and M. Schmaltz, Phys. Rev. D \textbf{90}, 054014 (2014) [arXiv:1408.1627]; 
 D. Aristizabal sierra, F. Staub and A. Vicente, Phys. Rev. D \textbf{92}, 015001 (2015) [arXiv:1503.06077];
Sofiane M. Boucenna, Jose W. F. Valle and A. Vicente, [arXiv:1503.07099];
F. Mahmoudi, S. Neshatpour, J. Virto, Eur. Phys. J. C \textbf{74} (2014) 2927, [arXiv:1401.2145];
A. Crivellin, G. D'Ambrasio, J. Heeck,  Phys. Rev. Lett. \textbf{114}, 151801 (2015) [arXiv:1501.00993];
A. Crivellin, G. D'Ambrasio, J. Heeck,  Phys. Rev. D \textbf{91}, 075006 (2015) [arXiv:1503.03477];
A. Crivellin, L. Hofer, J. Matias, U. Nierste, S. Pokorski, J. Rosiek, Phys. Rev. D {\bf 92}, 054013 (2015) [arXiv:1504.07928];
D. Becirevic, S. Fajfer, N. Kosnik, Phys. Rev. D {\bf 92}, 014016 (2015) [arXiv:1503.09024];
R. Alonso, B. Grinstein and J. Martin Camalich, Phys. Rev. Lett. \textbf{113}, 241802 (2014) [arXiv:1407.7044];
B. Gripaios, M. Nardecchia, S. A. Renner, JHEP \textbf{1505} (2015) 006, 
[arXiv:1412.1791]; A. Falkowski, M. Nardecchia, Robert Ziegler, [arXiv:1509.01249]; L. Calibbi, A. Crivellin, T. Ota, [arXiv:1506.02661];  S.~Descotes-Genon, J.~Matias and J.~Virto, 
  Phys.\ Rev.\ D {\bf 88}, 074002 (2013),
  [arXiv:1307.5683 [hep-ph]].
%%%%%%%%%%%%%%%%%%%%%%%%%%%%%%%%%%%%%%%%%%%%%%%%%%%%%%%%%%%%%%%%%%%%%%%%%%%%%%%%%%%%%

\bibitem{dyk} S. Meinel, D. van Dyk, [arXiv:1603.02974]. 

\bibitem{lhcb-lambda}
R. Aaij et al., [LHCb Collaboration], JHEP \textbf{06}, 115 (2015) [arXiv:1503.07138].


\bibitem{chen1}
Chaun-Hung Chen and C.Q. Geng, Phys. Rev. D \textbf{63}, 054005 (2001); Phys. Rev. D \textbf{63},  114024 (2001); Phys. Rev D \textbf{64},  074001 (2001).

\bibitem{Azizi} V. Bashity and K. Azizi, JHEP {\bf 0707}, 064 (2007) [arXiv:hep-ph/0702044]. 


\bibitem{mohanta3}
A. K. Giri and R. Mohanta, Eur. Phys. J. C\textbf{45}, 151-158 (2006), [arXiv:hep-ph/0510171]; Jour. Phys. G {\bf 31}, 1559 (2005);
R. Mohanta and A. K. Giri, Phys. Rev. D {\bf 82}, 094022 (2010)
[arXiv:1010.1152]


\bibitem{chen2}
Chaun-Hung Chen and C.Q. Geng, Phys. Lett. B {\bf 516}, 327 (2001) [arXiv:hep-ph/0101201]; C. S. Huang and H. J. Han, Phys. Rev. D {\bf 59}, 
114022 (1999); Erratum {\it ibid}, {\bf 61}, 039901 (2000).


\bibitem{lambda} 
W. Detmold, S. Meinel, Phys. Rev. D {\bf 93}, 074501 (2016) [arXiv:1602.01399];
 G. Kumar and N. Mahajan, [arXiv:1511.00935];  W. Detmold, C-J David Lin, S. Meinel and M. Wingate, Phys. Rev. D \textbf{87}, 074502 (2013); K. Azizi and N. Katirci Eur. Phys. J. A \textbf{48}, 73 (2012); T. M. Aliev, A. Ozpineci, M. Savci, Nucl. Phys. B \textbf{649}, 168-188 (2003), [arXiv:hep-ph/0202120]; T. M. Aliev, A. Ozpineci and M. Savci, Phys. Rev. D {\bf 65},
115002 (2002); {\it ibid} {\bf D} {\bf 67}, 035007 (2003); Nucl. Phys. B 
{\bf 709}, 115 (2005); T. M. Aliev, A. Ozpineci, M. Savci and C. Y\"uce, 
Phys. Lett. B {\bf 542}, 249 (2002); T. Gutsche, M. A. Ivanov, J. G. Korner, V. E. Lyubovitskij, P. Santorelli, Phys.Rev.D \textbf{87}, 074031 (2013), [arXiv:1301.3737];  P.~Boer, T.~Feldmann and D.~van Dyk,
  JHEP {\bf 1501}, 155 (2015), 
  [arXiv:1410.2115 [hep-ph]]; T.~Feldmann and M.~W.~Y.~Yip,
  Phys.\ Rev.\ D {\bf 85}, 014035 (2012), 
  Erratum: [Phys.\ Rev.\ D {\bf 86}, 079901 (2012)], 
  [arXiv:1111.1844 [hep-ph]].


\bibitem{LCSR}
Yu-ming Wang, Ying Li and Cai-Dian Lu, Eur. Phys. J.C \textbf{59}, 861 (2009), [arXiv:0804.0648].

\bibitem{Lattice-QCD}
W. Detmold, C. J. David Lin,  S. Meinel,  and M. Wingate, [arXiv: 1212.4827[hep-lat]].

\bibitem{Latt-QCD-2}
W. Detmold, S. Meinel, Phys. Rev. D \textbf{93}, 074501 (2016), [arXiv: 1602.01399[hep-lat]].




 \bibitem{georgi}
  H. Georgi and S. L. Glashow, Phys. Rev. Lett. \textbf{32}, 438 (1974).
  
  
   \bibitem{georgi2}
   H. Georgi, AIP Conf. Proc. \textbf{23} 575 (1975); H. Fritzsch and P. Minkowski, Annals Phys. \textbf{93}, 193 (1975); P. Langacker, Phys. Rep. \textbf{72},  185 (1981).
   
 \bibitem{pati} J. C. Pati and A. Salam, Phys. Rev. D \textbf{10}, 275 (1974).   
   
\bibitem{schrempp}
B. Schrempp and F. Shrempp, Phys. Lett.B \textbf{153}, 101 (1985); B. Gripaios, JHEP \textbf{1002}, 045 (2010) [arXiv:0910.1789].

  \bibitem{kaplan}
  D. B. Kaplan, Nucl. Phys. B \textbf{365}, 259 (1991).
  

 \bibitem{mohanta2}
 S. Sahoo and R. Mohanta, Phys. Rev. D \textbf{91}, 094019 (2015) [arXiv:1501.05193].

 \bibitem{mohanta1}
 R. Mohanta, Phys. Rev. D \textbf{89}, 014020 (2014) [arXiv:1310.0713].
 
 \bibitem{mohanta-Bs-mixing}
 
 S. Sahoo and R. Mohanta, Phys.Rev. D {\bf 93}, 034018 (2016) [arXiv:1507.02070].
 
  \bibitem{davidson}
   S. Davidson, D. C. Bailey and B. A. Campbell, Z. Phys.
C \textbf{61}, 613 (1994), hep-ph/9309310; I.
Dorsner, S. Fajfer, J. F. Kamenik, N. Kosnik, Phys. Lett. B
\textbf{682}, 67 (2009); [arXiv:0906.5585]; S. Fajfer, N. Kosnik, Phys. Rev. D \textbf{79}, 017502 (2009), [arXiv:0810.4858];
R. Benbrik, M. Chabab, G. Faisel,[ arXiv:1009.3886]
; A. V. Povarov, A. D. Smirnov, [arXiv:1010.5707]; J. P Saha, B. Misra and A. Kundu, Phys. Rev.D \textbf{81}, 095011 (2010) [arXiv:1003.1384]; 
I. Dorsner, J. Drobnak, S. Fajfer, J. F. Kamenik, N. Kosnik, JHEP \textbf{11}, 002 (2011) [arXiv: 1107.5393]; F. S. Queiroz, K.
Sinha, A. Strumia, [arXiv:1409.6301]; B. Allanach,A. Alves, F. S. Queiroz, K. Sinha, A. Strumia, [arXiv:1501.03494];
%%%%%%%%%%%%%%%%%%%%%%%%%%%%%%%%%%%%%%%%%%%%%%%%%%%%%%%%%%%%%%%%%%%%%%%%%%%%%%%%%%%%%%%
 Ivo de M. Varzielas, G. Hiller, JHEP 1506 (2005) 072, [arXiv:1503.01084];  
  S. Sahoo and R. Mohanta,
New Jour. Phys. {\bf 18}, 013032 (2016) [arXiv:1509.06248]; M. Bauer and M. Neubert, [arXiv:1511.01900];
S. Fajfer and N. Kosnik, [arXiv:1511.06024].  
%%%%%%%%%%%%%%%%%%%%%%%%%%%%%%%%%%%%%%%%%%%%%%%%%%%%%%%%%%%%%%%%%%%%%%%%%%%%%%%%%%%%%%%%%%%%
\bibitem{arnold}
 J. M. Arnold, B. Fornal and M. B. Wise, Phys. Rev. D \textbf{88}, 035009 (2013), [arXiv:1304.6119].
 
\bibitem{kosnik}
  N. Kosnik, Phys. Rev. D \textbf{86}, 055004 (2012), [arXiv:1206.2970].
 

\bibitem{leptoquark}
D. Aristizabal Sierra, M. Hirsch, S. G. Kovalenko, Phys. Rev. D \textbf{77}, 055011 (2008), [arXiv:0710.5699]; 
K.S. Babu, J. Julio, Nucl. Phys. B \textbf{841}, 130 (2010), [arXiv:1006.1092]; S. Davidson, S. Descotes-Genon, 
JHEP \textbf{1011}, 073 (2010), [arXiv:1009.1998]; S. Fajfer, J. F. Kamenik, I. Nisandzic, J. Zupan, Phys. Rev. Lett. \textbf{109}, 161801, 
(2012), [arXiv:1206.1872]; K. Cheung, W.-Y. Keung, P.-Y. Tseng, [arXiv:1508.01897]; D. A. Camargo, [arXiv:1509.04263]; 
S. Baek, K. Nishiwaki, [arXiv:1509.07410]; Heinrich Pas, Erik Schumacher, Phys. Rev. D \textbf{92}, 114025 (2015)  [arXiv:1510.08757];  S.~M.~Boucenna, A.~Celis, J.~Fuentes-Martin, A.~Vicente and J.~Virto,
  Phys.\ Lett.\ B {\bf 760}, 214 (2016), 
  [arXiv:1604.03088 [hep-ph]].

\bibitem{mohanta4}
 S. Sahoo and R. Mohanta, Phys. Rev. D {\bf 93}, 114001 (2016) [arXiv:1512.04657].
 
 \bibitem{wang}
Shuai-wei Wang, Ya-dong Yang, [arXiv: 1608.03662[hep-ph]].
 
\bibitem{buras} G.\ Buchalla, A.J.\ Buras, M.\ Lautenbacher, Rev.\
Mod.\ Phys.\ {\bf 68}, 1125 (1996).



\bibitem{wilson} 
W. Altmannshofer, P. Ball, A. Bharucha, A. J. Buras, D. M. Straub and M. Wick,  	JHEP \textbf{0901}, 019 (2009), [arXiv:0811.1214].

\bibitem{buras2}
A. J. Buras and M. Munz, Phys. Rev. D \textbf{52} (1995) 186; M. Misiak, Nucl. Phys. B 393 (1993) 23; Erratum, \textit{ibid}. B \textbf{439},   461
(1995).



\bibitem{res} C. S. Lim, T. Morozumi and A. I. Sanda, Phys. Lett. B. 
{\bf 218}, 343 (1989); N. G. Deshpande, J. Trampetic and K. Ponose, Phys. 
Rev. D {\bf 39}, 1461 (1989);
P. J. O'Donnell and H. K.K. Tung, Phys. Rev. D {\bf 43}, 2067 (1991);
P. J. O'Donnell, M. Sutherland  and H. K.K. Tung, Phys. Rev. D {\bf 46}, 
4091 (1992); F. Kr\"uger and L. M. Sehgal, Phys. Lett. B {\bf 380}, 199
(1996).


\bibitem{caso}
 A. Ali, P. Ball, L. T. Handoko, and G. Hiller, Phys. Rev. D\textbf{61},  074024 (2000).
 
\bibitem{mannel} T. Mannel, W. Roberts and Z. Ryzak, Nucl. Phys. B {\bf 355},
38 (1991); T. Mannel and S. Recksiegel J. Phys. \textbf{G24}, 979 (1998).



 \bibitem{bobeth1}
C. Bobeth, M. Gorbahn, T. Hermann, M. Misiak, E. Stamou, M. Steinhauser, Phys. Rev.Lett.
\textbf{112}, 101801 (2014) [arXiv:1311.0903].

\bibitem{cms}
S. Chatrchyan et al., [CMS Collaboration], Phys. Rev. Lett. \textbf{111}, 101805 (2013),
[arXiv:1307.5025].

\bibitem{lhcb6}
 R. Aaij et al., [LHCb Collaboration], Phys. Rev. Lett. \textbf{111}, 101805 (2013), [arXiv:1307.5024].
 
\bibitem{lhcb7}
V. Khachatryan et al. [CMS Collaboration] and I. Bediaga
et al. [LHCb Collaboration], Nature (London) {\bf 522}, 68
(2015).

\bibitem{lim}
T. Inami and C. S. Lim, Prog. Theor. Phys.
\textbf{65}, 297 (1981); [Erratum-ibid. \textbf{65}, 1772 (1981)].

\bibitem{ckmfitter}
J. Charles {\it et al.}, Phys. Rev. D \textbf{91}, 073007 (2015), [arXiv:1501.05013].

\bibitem{lfv}
 I. Ilakovac and A. Pilaftsis, Nucl. Phys. B {\bf 437}, 491 (1995) [hep-ph/9403398];
R. Barbieri, L. H. Hall and A. Strumia, Nucl. Phys. B {\bf 445}, 219 (1995) [hep-ph/9501334];
J. Hisano, T. Moroi, K. Tobe and M. Yamaguchi, Phys. Rev. D. {\bf 53}, 2442 (1996) [hep-ph/9510309];
J. R. Ellis, J. Hisano, M. Raidal and Y. Shimizu, Phys. Rev. D {\bf 66}, 115013 (2002) [hep-ph/0206110];
A. Brignole and A. Rossi, Nucl. Phys. B {\bf 701}, 3 (2004) [hep-ph/0404211];
A. Masiero, S. Profumo, S. Vempati and C. E. Yaguna, JHEP {\bf 0403} 046 (2004) [hep-ph/0401138];
A. Arganda and M. J. Herrero, Phys. Rev. D {\bf 73}, 055003 (2006) [hep-ph/0510405];
A. Antusch, E. Erganda, M. J. Herrero and A. M. Teixeira, JHEP {\bf 0611} 090 (2006) [hep-ph/0607263];
P. Paradisi, JHEP {\bf 0510}, 006 (2005) [hep-ph/0505046]; JHEP {\bf 0602} 050 (2006)
[hep-ph/0508054]; JHEP {\bf 0608}, 047 (2006) [hep-ph/0611100];
A. G. Akeroyd, M. Aoki and Y. Okada, Phys. Rev. D {\bf 76}, 013004 (2007) [hep-ph/0610344];
 B. M. Dassinger, T. Feldmann, T. Mannel, S. Turczyk, JHEP {\bf 0710}, 039 (2007)
[arXiv:0707.0988]; R. Mohanta, Euro Phys. J. C {\bf 71}, 1625 (2011) [arXiv:1011.4184];
R. Alonso, B. Grinstein and J. M. Camalich, [arXiv: 1505.05164]; Chao-Jung Lee and
J. Tandean, JHEP {\bf 08}, 123 (2015), [arXiv: 1505.04692]; W. Altmannshofer and I. Yavin,
Phys Rev D. {\bf 92}, 075022 (2015) [arXiv:1508.07009].




\end{thebibliography}
\end{document}